\documentclass[preprint,10pt]{aastex}

\usepackage{epsfig}
\usepackage{graphicx}
\usepackage{amsmath,amssymb}
\usepackage{natbib}
\usepackage{subfigure}
\usepackage{mathrsfs}

\begin{document}

\title{Strong field effects on emission line profiles: Kerr black holes and warped accretion disks}

\author{Yan Wang and Xiang-Dong Li}
\affil{Department of Astronomy, Nanjing University, Nanjing 210093,
  China}
\affil{Key Laboratory of Modern Astronomy and Astrophysics (Nanjing University),
Ministry of Education, Nanjing 210093, China}
%

%

\begin{abstract}
 If an accretion disk around a black hole is illuminated by hard X-rays from
 non-thermal coronae, fluorescent iron lines will be emitted from the inner region
 of the accretion disk. The emission line profiles will show a variety of strong
 field effects, which may be used as a probe of the spin parameter of the black
 hole and the structure of the accretion disk. In this paper we generalize the previous
 relativistic line profile models by including both the black hole spinning effects and the
 non-axisymmetries of warped accretion disks. Our results show different features
 from the conventional calculations for either a flat disk around a Kerr black hole
 or a warped disk around a Schwarzschild black hole by presenting, at the same time, multiple
 peaks, rather long red tails and time variations of line profiles with the precession
 of the disk. We show disk images as seen by a distant observer, which are distorted by the 
 strong gravity. Although we are primarily concerned with the iron K-shell lines in this 
 paper, the calculation is general and is valid for any emission lines produced from a warped
 accretion disk around a black hole.

\end{abstract}

\maketitle

\section{Introduction}\label{sec:intro}

 Accretion power serves as an important source of energy in astrophysics,
 especially in galactic X-ray binaries (XRBs) and active galactic nuclei (AGNs).
 In these systems, black holes (BHs) are the possible candidates for the central
 accreting compact objects. Accretion onto both stellar-mass black holes in XRBs
 and supermassive black holes (${10}^{6}-{10}^{9}M_{\odot}$) in AGNs usually
 proceeds through an accretion disk, which may be truncated at the inner most
 stable circular orbit (ISCO) because of the general relativistic effect of black hole.
 Due to viscosity, plasma elements in the disk gradually spiral inwards to the
 black hole while losing their angular momenta until reaching the ISCO; they then plunge
 into the black hole, if there is no external force acting on them.

 The quasi-thermal spectra of the black hole accretion disks are relatively cold,
 which peak at optical/UV band in AGNs and soft X-ray band in XRBs \citep{Frankbook2002}.
 If point-like hot coronae, above and below the center of the accretion disk (as in
 the lamp-post model \citep{1991A&A...247...25M}), illuminate the inner region of the disk by hard
 X-ray photons yielded via thermal Comptonization, iron atoms photoelectrically absorb
 the incident hard X-ray photons, then emit iron K-shell photons (centered at $6.40-6.97$ keV,
 depending on the ionization state) through the fluorescent mechanism. Iron K-shell
 lines may be the strongest emission lines in the X-ray observations of
 XRBs (\citet{1985MNRAS.216P..65B}, \citet{2002ApJ...578..348M} for Cygnus X-1;
 \citet{2002ApJ...570L..69M}, \citet{2004MNRAS.351..466M}, \cite{2005MNRAS.360..763R}
 for XTE J1650-500) and AGNs (\citet{1995Natur.375..659T}, \cite{2001MNRAS.328L..27W},
 \citet{2006ApJ...652.1028B}, \citet{2007PASJ...59S.315M} for the Seyfert-1 MCG-6-30-15,
 the most important source to date). Observations suggest that some two-horn broadened
 iron K-shell lines originate from the innermost area of the accretion disk. The iron
 lines could be a powerful tool to study the strong gravity within a few gravitational
 radius $r_g=GM/c^2$ and a probe of the black hole spin parameter
 (see reviews by \citet{2003PhR...377..389R} and \citet{2007ARA&A..45..441M}).

 \citet{1989MNRAS.238..729F} modeled the relativistic line profile from a standard thin
 disk (see \cite{1973A&A....24..337S} for Newtonian disk model; \citet{1973blho.conf..343N}
 and \citet{1974ApJ...191..499P} for general relativistic effects) orbiting around
 a Schwarzschild black hole. \citet{1991ApJ...376...90L} and \citet{1997ApJ...475...57B}
 calculated the line profile from a standard thin disk orbiting around an extreme
 Kerr black hole. \citet{1997PASJ...49..159F} calculated the line profile from
 a standard thin disk orbiting around a Kerr black hole with spin parameter
 $a=cJ/GM$ (where $J$ is the angular momentum and $M$ is the mass
 of the black hole) as a free parameter, in their calculation the complication of
 computing azimuthal angle $\phi$ has been avoided due to the axisymmetry of
 both the Kerr metric and the accretion disk. Thereafter, the line profiles from accretion
 disks with different structures have been investigated: for a warped accretion disk orbiting around
 a Schwarzschild black hole \citep{2000MNRAS.317..880H,2008MNRAS.389..213W}, and for a flat 
 accretion disk with spiral pattens orbiting around a nonrotating black hole
 \citep{2002MNRAS.332L...1H} or a rotating black hole \citep{2001PASJ...53..189K}. 
 In these calculations, the accretion disks are optically thick
 and geometrically thin; furthermore it was assumed that the plasma elements within ISCO
 contribute little to the line profiles, because they are hot and almost fully ionized
 \citep{1998MNRAS.300L..11Y,2006ApJ...652.1028B}. In contrast with the thin disk,  
 \citet{2007MNRAS.378..841W} explored the role of disk self-shadowing effect due 
 to significant geometrical thickness. \citet{2007A&A...474...55F} investigated the 
 influence of geometrical thickness of relativistic accretion torus on the line profiles.

 It is interesting to model the relativistic iron K-shell lines emerging
 from the warped disks orbiting around Kerr black holes.  \citet{1975ApJ...195L..65B}
 showed that for Kerr black holes the dragging of inertial frames can lead to
 coupling between the spin of the black hole and the orbital angular momentum of the
 disk, and that the component of the viscous torque, which is in the disk plane and perpendicular
 to the line of nodes, will tend to force the inner part of the disk to lie on the
 equatorial plane of the Kerr black hole. Nevertheless, misalignment of the inner part
 of the disk may take place even in the presence of dissipation. \citet{1997A&A...324..829D}
 and \citet{1997MNRAS.285..394I} showed that the misaligned disk adopts a steady warped
 shape in which the tilt angle is an oscillatory function of radius. \citet{2002MNRAS.337..706L}
 considered the time evolution of the warped disk under the condition that the warping
 propagates in a wave like manner, and concluded that, in low-viscosity disk around a Kerr
 black hole, the inner parts of the disk are not necessarily aligned with the black hole.
 Warping may be induced by non-axisymmetric radiation pressure \citep{1996MNRAS.281..357P,
 1998ApJ...504...77M} or tidal interaction \citep{1993A&A...274..291T,1996MNRAS.282..597L}.
 In observations, evidence of warped disks in XRB includes the 35-days periodic variation
 of X-ray flux of Her X-1 \citep{1972ApJ...174L.143T,1973Natur.246...87K} and the precessing
 relativistic jet in SS 433 \citep{1984ARA&A..22..507M}; for AGNs disk warping can explain
 the radial dependence of declination between high-velocity maser and systemic maser in
 NGC 4258 \citep{1995Natur.373..127M,2008MNRAS.387..830M} and the misalignment between the
 angular momentum of the radio jets and the host galactic disks \citep{2000ApJ...537..152K,
 2002ApJ...575..150S}.

 The motivation of this work is multiple. Firstly, the iron K-shell
 lines carry the information about the spacetime around a rotating back hole; fitting line
 profiles from the observational data with better spectral resolution and signal-noise
 ratio by more realistic disk line models may provide more reliable estimations of the
 spin parameter of the black hole \citep{2010A&A...521A..15A}, and distinguish the warping
 effects on line profiles from others. Secondly, the disk lines may be used to
 diagnose the geometry and kinematics of the disk itself, and provide observational signatures
 for the existence and characteristics (magnitude, twist-free or twisted) of the warping.

 The paper is organized as follows. In Sec.~\ref{sec:geodesic} we
 review the properties of photon trajectories in the Kerr spacetime,
 and summarize the relevant equations needed in our calculation.
 Next, we describe in Sec.~\ref{sec:struc}.1 the geometric formalism
 for the warped disk and in Sec.~\ref{sec:struc}.2 the kinematic
 structure. These considerations are employed in Sec.~\ref{sec:profile},
 where we show our main results, including the emission line profiles
 and the images of the warped disk around a Kerr black hole observed
 from a distant place. We summarize the main results and discuss their possible
 implications in Sec.~\ref{sec:conc}. To allow the main ideas of the
 paper to be as clear as possible, several sets of details have been
 relegated to appendix A, in which we give the detailed procedure for
 deriving the expression of the ratio $g$ for the warped disk. The
 relativistic calculations presented in this paper use the notational
 conventions of the text by \citet{MTW}; in particular we use $c=G=1$.

\section{Photon trajectories in the Kerr spacetime}\label{sec:geodesic}

 When calculating the relativistic line profiles, one needs to trace the
 trajectories of the photons in the Kerr spacetime. In the standard
 Boyer-Lindquist coordinates, the Kerr metric can be written as
\begin{equation}\label{eq:kerr}
ds^2=-\Sigma\Delta A^{-1}dt^2+\sin^2\theta A\Sigma^{-1}\left(d\phi-\omega
dt\right)^2+\Sigma\Delta^{-1} dr^2+\Sigma d\theta^2\,,
\end{equation}
 where
\begin{equation}\label{eq:kerr1}
\Sigma=r^2+a^2\cos^2\theta\,,
\end{equation}
\begin{equation}\label{eq:kerr2}
\Delta=r^2+a^2-2Mr\,,
\end{equation}
\begin{equation}\label{eq:kerr3}
A=\left(r^2+a^2\right)^2-a^2\sin^2\theta\Delta=\Sigma\left(r^2+a^2\right)
+2Ma^2r\sin^2\theta\,,
\end{equation}
\begin{equation}\label{eq:kerr4}
\omega=2aMrA^{-1}\,.
\end{equation}
 This metric contains two parameters, the mass of the black hole $M$ and the
 specific angular momentum or the spin parameter $a=J/M$.
 The greater root $r_{+}=M+\sqrt{M^{2}-a^{2}}$ of $\Delta=0$ corresponds
 to the event horizon which is an one-way membrane in the Kerr spacetime.
 We notice that the right-hand-sides of Eqs.~\ref{eq:kerr1}--\ref{eq:kerr4} 
 are always positive outside the event horizon. The dragging of inertial 
 frame forbids any observer remaining static within the static limit
 $r_{\rm sl}(\theta)=M+\sqrt{M^{2}-a^{2}\cos^2\theta}$.

 Given a spacetime background, the motion of a photon obeys the null geodesic
 equation $\nabla_{\mathbf{p}}\cdot \mathbf{p}=0 $, where $ \mathbf{p} $ is
 the 4-momentum of the photon and $\nabla_{\mathbf{p}} $ is the directional
 derivative operator along $\mathbf{p}$\,. This is a set of second-order ordinary
 differential equation, which can be solved in principle by the Runge-Kutta
 method with given initial position and momentum of the photon. Alternatively,
 in Kerr spacetime one can take advantage of its axisymmetric property (the corresponding
 constants of motion include the energy of the photon $E$, the angular momentum
 along the spin axis $L_z$, and the norm of the 4-momentum $\delta$), and solve
 the trajectory of photon in the equatorial plane by the Lagrangian approach.
 For the general null geodesics in the Kerr spacetime, \citet{1968PhRv..174.1559C}
 demonstrated the separability of the Hamilton-Jacobi equation and discovered the
 existence the fourth constant of motion (i.e. Carter constant $\mathscr{Q}$).
 To obtain the equations governing the motion of the photons, one can take the
 partial derivatives of the Hamilton's principle function
\begin{equation}\label{eq:geokerr1}
S=\frac{1}{2}\delta\lambda-Et+L_{z}\phi+\int^r\frac{\sqrt{R(r)}}{\Delta} dr
+\int^{\theta}\sqrt{\Theta(\theta)}d\theta \,,
\end{equation}
 with respect to $E$, $L_z$, $\delta$, and $\mathscr{Q}$ then set them to zero.
 After some manipulations (see \citet{1983mtbh.book.....C} for details), one can
 find that the equation governing the motion of the photon ($\delta=0$) in $r$
 and $\theta$ is
\begin{equation}\label{eq:geokerr1}
\int^r\frac{dr}{\sqrt{R(r,\xi,\eta)}}=\int^{\theta}\frac{d\theta}{\sqrt{\Theta(\theta,\xi,\eta)}}=P\,,
\end{equation}
 where
\begin{equation}\label{eq:geokerr2}
R(r,\xi,\eta)=r^4+\left(a^2-\xi^2-\eta\right)r^2+2M\left(\eta+(\xi-a)^2\right)r-a^2\eta\,,
\end{equation}
\begin{equation}\label{eq:geokerr3}
\Theta(\theta,\xi,\eta)=\eta+a^2\cos^2\theta-\xi ^2\cot ^2\theta\,,
\end{equation}
 and $P$ is the parameter along the photon trajectory. Here we have defined two dimensionless 
 impact parameters $\xi=L_{z}/E$ and $\eta=\mathscr{Q}/E^{2}$. The equations of motion
 for other coordinates (including the affine parameter $\lambda$) can be expressed
 as combinations of the integrals about $r$ and $\theta$
\begin{equation}\label{eq:geokerrlambda}
\lambda=\int^r\frac{r^2dr}{\sqrt{R(r,\xi,\eta)}}+\int^{\theta}\frac{a^2\cos ^2\theta
d\theta}{\sqrt{\Theta(\theta,\xi,\eta)}}\,,
\end{equation}
\begin{equation}\label{eq:geokerrt}
t=E\lambda+2M\int^r r\left(Er^2-aL_z+a^2E\right)\frac{dr}{\Delta\sqrt{R(r,\xi,\eta)}}\,,
\end{equation}
\begin{equation}\label{eq:geokerrphi}
\phi=a\int^r\left(( r^2+a^2 )E-aL_z\right)\frac{dr}{\Delta\sqrt{R(r,\xi,\eta)}}+
\int^{\theta}\left(L_z\csc^2\theta-aE\right)\frac{d\theta}{\sqrt{\Theta(\theta,\xi,\eta)}}\,.
\end{equation}
 The upper and lower limits in the integrals
 about $r$ and $\theta$ correspond to the turning points of the photon
 trajectories which are the roots of quartic Eqs.~\ref{eq:geokerr2} and
 \ref{eq:geokerr3} with different choices of the impact parameters. The
 limits are not written explicitly here due to the complexity (see Appendix A
 in \citet{1994ApJ...421...46R} and Tables in \citet{2009ApJ...696.1616D}
 for details). Eqs.~\ref{eq:geokerr1}, \ref{eq:geokerrlambda}, \ref{eq:geokerrt},
 and \ref{eq:geokerrphi} are elliptical integrals, which can be reduced to
 the Carlson's standard forms, then be evaluated numerically by the subroutines
 provided in \citet{1992nrfa.book.....P}. In the calculation of photon
 trajectories, we use the efficient subroutine \texttt{geokerr} created
 by \citet{2009ApJ...696.1616D} which has integrated the necessary elements
 described above, to calculate all geodesic coordinates semi-analytically.
 In addition we shoot photons from the observer's plane of the sky backwards
 towards the central black hole and search for the intersections of the
 photons with the disk. This is a more efficient way than tracing the photons
 from the disk to the observer in our application.


\section{Geometric and kinematic structure of the warped accretion disk}\label{sec:struc}

\subsection{Geometric structure of the disk}


\begin{figure}[h]
\begin{center}
\includegraphics[width=.5\textwidth]{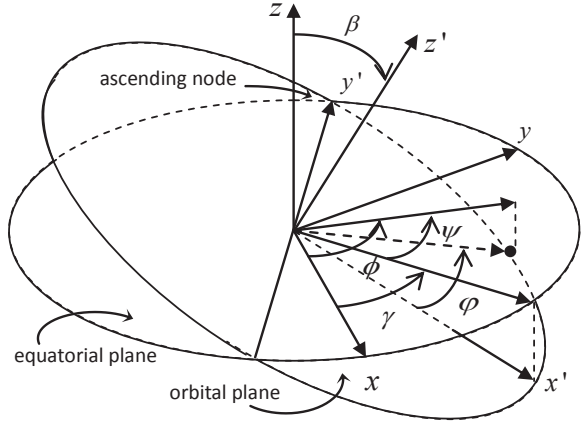}
\caption{The geometry of the equatorial plane of the rotating black hole and the
 orbital plane of the plasma particle at $r$. The particle orbits around the
 central black hole along the circular ring in the orbital plane. The concentric
 rings with different $\beta$ and $\gamma$ compose of the warped accretion disk. }
\label{fig:pic2}
\end{center}
\end{figure}

 In our calculations, we use a series of concentric rings with increasing radii
 to describe the shape of the geometrically thin warped disk. The spherical 
 ($r$,~$\theta$,~$\phi$) and cylindrical coordinate system ($\rho$,~$\phi$,~$h$) 
 are both appeared in our calculations, and they can be transformed from each other 
 by 
\begin{eqnarray}
 h &=& r\sin\beta\cos\varphi \nonumber \\
   &=& \rho\tan\beta\cos\psi\,,
\end{eqnarray}
 and
\begin{eqnarray}
 \rho &=& r\left(1-\sin^2\beta\cos^2\varphi\right)^{1/2} \nonumber \\
      &=& r\left(1-\frac{\sin^2\beta}{1+\tan^2\psi\cos^2\beta}\right)\,,
\end{eqnarray}
 where we use $\rho\cos\psi=r\cos\varphi\cos\beta$,
 $\tan\varphi=\tan\psi\tan\beta$ and $\psi=\phi-\gamma$ in the derivation above. 
 The formalism of the warped disk has been discussed by \citet{1993A&A...274..291T,
 2000MNRAS.317..880H,2008MNRAS.389..213W}.
 Here we adopt the function expression of the two Eulerian angles used
 by \citet{2008MNRAS.389..213W}, and the warping of disk is in an oscillation shape:

\begin{equation}\label{eq:twist}
\gamma=\gamma_0+n_{1}\exp\left({n_{2}\frac{r_{\text{in}}-r}{r_{\text{out}}-r_{\text{in}}}}\right)  \,,
\end{equation}
\begin{equation}\label{eq:tilt}
\beta=n_3\sin\left[\frac{\pi}{2}\left(\frac{r-r_{\text{in}}}{r_{\text{out}}-r_{\text{in}}}\right)\right] \,.
\end{equation}
 We further specify that the inner radius of the disk $r_{\text{in}}$ is always 
 fixed at the ISCO, and the outer radius $r_{\text{out}}$ at $50~r_g$. 
 The parameters $n_1$, $n_2$, $n_3$ are used to describe the magnitude 
 of warping. $\gamma_0$ measures the azimuthal viewing angle
 of the observer relative to the warped disk. Choosing different values
 for $\gamma_0$ is equivalent to changing the observational time during the
 disk precession. Therefore we can study the time variations of the line
 profiles and the disk images.

\subsection{Kinematic structure of the disk} \label{sec:struc2}

 The kinematic structure of the warped disk carries the information about
 the motion of emitting particles in the vicinity of the Kerr black hole.
 It mainly contributes to the Doppler shifts (the only contributor to blue-shift)
 in the line profiles. We consider that each plasma element follows a
 quasi-Keplerian orbit in a concentric ring, which means the radial velocity
 is much less than the transverse velocity as in the standard thin disks,
 and the angular velocity of the spatially circular orbit can be written as
 \begin{equation}\label{eq:angularV}
 \Omega=(a+\epsilon\sqrt{r^3/M})^{-1}\,,
 \end{equation}
 where $\epsilon=1$ corresponds to the prograde orbit and $\epsilon=-1$
 the retrograde orbit with respect to the spinning of the black hole. This
 angular velocity is modified by the relativistic effect of the black hole by the
 inclusion of an additional spin parameter $a$. Strictly speaking, the circular orbit
 exists only in the equatorial plane. However we assume that the tiled angle is small so
 that Eq.~\ref{eq:angularV} is a good approximation for the warped disk. For the clarity
 in this paper, we simply list the 4-velocity of particles below and relegate the
 detailed derivation to Appendix A.
\begin{equation}\label{eq:u0f}
u^{0}=\left(\Sigma\Delta A^{-1}-\sin^2\theta A\Sigma^{-1}\left(\dot{\phi}-\omega\right)^2
-\Sigma\dot{\theta}^2\right)^{-1/2} \,,
\end{equation}
\begin{equation}\label{eq:urf}
u^{r}=0 \,,
\end{equation}
\begin{equation}\label{eq:uthetaf}
u^{\theta}=\dot{\theta}u^{0}=\Omega u^0\left(-\cos\psi\cos\beta\sin\varphi\cos\theta
 +\sin\psi\cos\varphi\cos\theta+\sin\beta\sin\varphi\sin\theta\right)  \,,
\end{equation}
\begin{equation}\label{eq:uphif}
u^{\phi}=\dot{\phi}u^{0}=\frac{\Omega u^0}{\sin\theta}\left(\sin\psi\cos\beta\sin\varphi+\cos\psi\cos\varphi\right)  \,.
\end{equation}

\section{Emission line profiles and warped disk images}\label{sec:profile}

\subsection{Emission line profiles}

 The emission line profile produced by an accretion disk around a black hole is influenced
 by Doppler shift, gravitational red-shift, beaming effect, and gravitational lensing effect.
 The changing of photon energy along a bundle of photon trajectories (geodesic congruence)
 can be characterized by the ratio $g$ of the energy $E_{\text{obs}}$ measured by a local
 inertial (rest) observer at asymptotic infinity to the emitted energy $E_{\text{em}}$
 measured by an observer comoving with the plasma element on the accretion disk,
 \begin{equation}
 g=\frac{E_{\text{obs}}}{E_{\text{em}}}
 =\frac{(u^{\mu}p_{\mu})_{\text{obs}}}{(u^{\mu}p_{\mu})_{\text{em}}}=(1+z)^{-1} ~\,,
 \end{equation}
 where $z\equiv(\lambda_{\text{obs}}-\lambda_{\text{em}})/\lambda_{\text{em}}$ is the red-shift
 as usually defined, $u^{\mu}$ in the denominator and numerator are the four-velocity of the
 emitting plasma element and the observer at infinity respectively, and $p_{\mu}$ is the
 four-momentum of the photon which propagates from the plasma element to the observer along a
 null geodesic. The total observed flux $F_{\text{obs}}$ is determined by integrating the specific
 flux $\text{d}F_{\text{obs}}$ over all the plasma elements on the accreting disk. For the line profile,
 the specific flux $\text{d}F_{\text{obs}}(E_{\text{obs}})$ at observed energy $E_{\text{obs}}$ can be
 expressed as:
 \begin{equation}
 \text{d}F_{\text{obs}}(E_{\text{obs}})=I_{\text{obs}}(E_{\text{obs}})\text{d}\Omega_{\text{obs}} ~\,,
 \end{equation}
 where $I_{\text{obs}}$ is the observed specific intensity and $\text{d}\Omega_{\text{obs}}$ is the
 observed solid angle subtended by each plasma element. Using Liouville's theorem
 \begin{equation}
 I_{\text{obs}}/\nu_{\text{obs}}^{3}=I_{\text{em}}/\nu_{\text{em}}^{3}
 \end{equation}
 we find
 \begin{equation}
 I_{\text{obs}}=I_{\text{em}}\nu_{\text{obs}}^{3}/\nu_{\text{em}}^{3}=I_{\text{em}}g^{3}
               =I_{\text{em}}(1+z)^{-3} ~\,.
 \end{equation}
 And the observed solid angle $\text{d}\Omega_{\text{obs}}$ is naturally the size of the pixels
 in the observer's screen. The detailed derivation of the factor $g$ for the warped disk
 is shown in Appendix~\ref{sec:appA}. We ignore the intrinsic line width and assume
 $I_{\text{em}}=\varepsilon\delta(E_{\text{em}}-E_{0})$, where $E_{0}$ is the rest energy
 of photons, $\varepsilon\sim(r/M)^{q}$ is the surface emissivity, and $q$ is the
 emissivity index \citep{1989MNRAS.238..729F}.

 In the actual computation, we trace every null geodesic from the observer's screen
 backwards towards the central black hole and search for the interactions of the
 trajectories with the surface of the disk. The shadowing from the observer due to the
 warping of the disk is taken into account; we ignore the shadowing from the
 coronae, since in our calculation we assume that the scale height of the coronae is large
 compared to the size of the disk of our interest, therefore this shadowing may not
 be important. We calculate the observed specific flux $\text{d}F_{\text{obs}}$ from each
 plasma element, and add its contribution into a energy bin according to the observed
 energy of photons ($E_{\text{obs}}=gE_{\text{em}}$). Then we can obtain a line profile
 by plotting count number ($\text{d}F_{\text{bin}}/E_{\text{bin}}$) versus $E_{\text{bin}}$.
 The number of bins we choose determines the spectrum resolution. We assume an optically
 thick disk, thus we consider only the direct photons and neglect the photons which would
 circular around the black hole and strike the disk.

\subsubsection{Changing of the spin parameter $a$}

 The overall properties of the line profiles can be understood by carefully examining the
 contribution of the specific flux from each of the narrow rings of the disk, they are
 determined by the spin parameter, the warping parameters, the emissivity index, shadowing
 and orientation of the disk relative to a distant observer (both the inclination angle and
 azimuthal angle).

\begin{figure}[h]
\begin{center}
\includegraphics[width=1.0\textwidth]{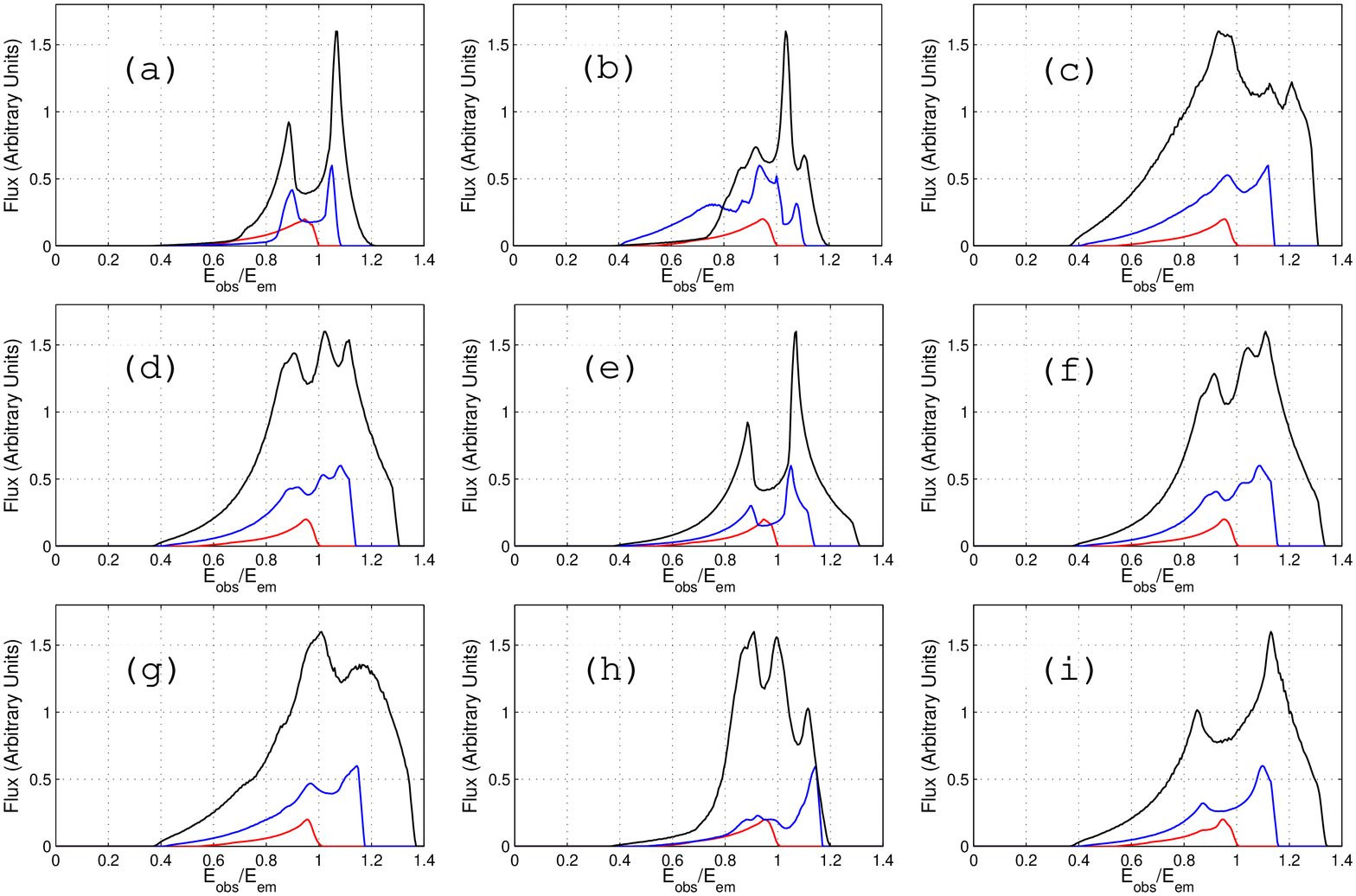}
\caption{Line profiles from a warped disk with $n_1=0$, $n_2=1$, $n_3=0.95$, $r_{in}=4.23~r_g$
 and $r_{out}=50~r_g$. The spin parameter $a$ of the Kerr black hole is $0.5M$. The emissivity index
 $q$ is taken to be $-2$. The horizontal axis is the $g$ factor, i.e. the observed photon energy
 per unit iron K-shell photon energy; the vertical axis is flux in arbitrary units. In each panel, three
 colored line profiles represent inclination angles $\theta$ of $10^{\circ}$ (red),
 $50^{\circ}$ (blue), and $70^{\circ}$ (black) measured from the spin axis of the black hole
 respectively. Panels (a)-(h) contain the line profiles seen from different azimuthal angles $\phi$ of
 $0^{\circ}$, $45^{\circ}$, $90^{\circ}$, $135^{\circ}$, $180^{\circ}$, $225^{\circ}$, $270^{\circ}$ and
 $315^{\circ}$. Panel (a) shows the $\phi=0^{\circ}$ case which is defined as the observer viewing
 from the direction furthest from the lowest point of the disk. Panel (i) shows the line profile
 from a standard flat disk for comparison.
 }
\label{fig:dell_kerr5SF1357}
\end{center}
\end{figure}

 Figure~\ref{fig:dell_kerr5SF1357} shows the line profiles from the warped accretion
 disk with the warping parameters $n_1=0$, $n_2=1$, and $n_3=0.95$. The inner radius
 $r_{\text{in}}$ of the prograde disk ($\epsilon=1$) locates at the innermost stable circular orbit (ISCO),
 which is $4.23~r_g$ for the black hole spin parameter $a=0.5M$ (maybe a good estimate
 for the supermassive black hole in Sgr A*). The outer radius
 is chosen as $r_{\text{out}}=50~r_g$. From Eq.~\ref{eq:tilt} we know that the tilt angles
 are $0^{\circ}$ and $54^{\circ}$ for the narrow rings of the disk at the inner radius
 and the outer radius respectively. This warped disk is twist-free which means that the
 twist angle $\gamma$ does not depend on the radius ($n_1=0$). The emissivity index $q$ is taken
 to be $-2$. The horizontal axis is the $g$ factor, i.e. the observed photon energy per
 unit iron K-shell photon energy, while the vertical axis gives flux in arbitrary units at
 different observed photon energies. In each panel, there are three colored line profiles
 representing the inclination angles $\theta$ of $10^{\circ}$ (red), $50^{\circ}$ (blue),
 and $70^{\circ}$ (black) measured from the spin axis of the black hole respectively.
 Panels (a)-(h) contain the line profiles seen from different azimuthal viewing angles,
 for the retrograde precession of the warped disk about the spin axis of the black hole
 from $0^{\circ}$ to $315^{\circ}$ with uniform angular intervals of $45^{\circ}$.
 Panel (a) shows the $\phi=0^{\circ}$ case, which is defined as the observer
 viewing from the direction furthest from the lowest point of the disk. Panel (i)
 shows the line profiles from the standard flat disk with the same inner
 and outer radius for comparison.

\begin{figure}
\begin{center}
\begin{tabular}{ccc}
\epsfig{file=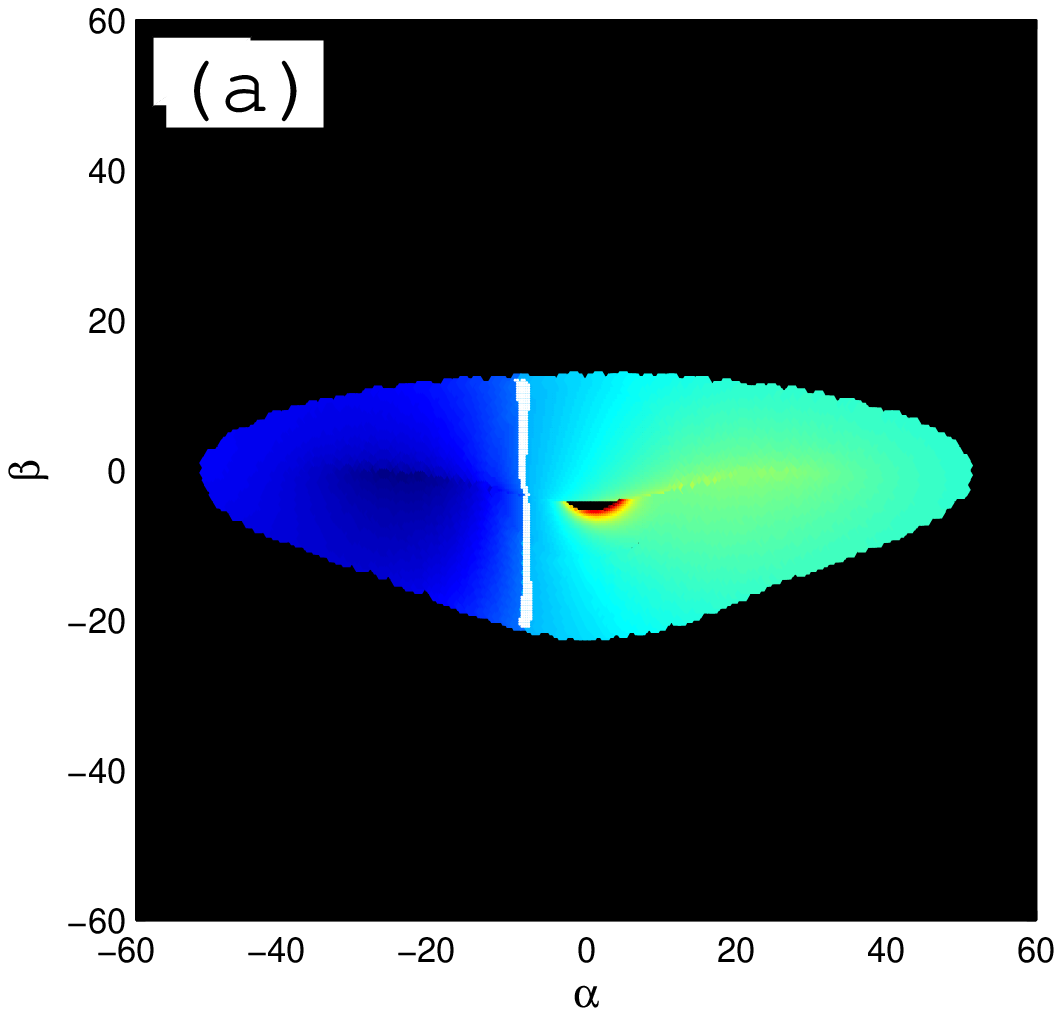,width=0.3\linewidth} &
\epsfig{file=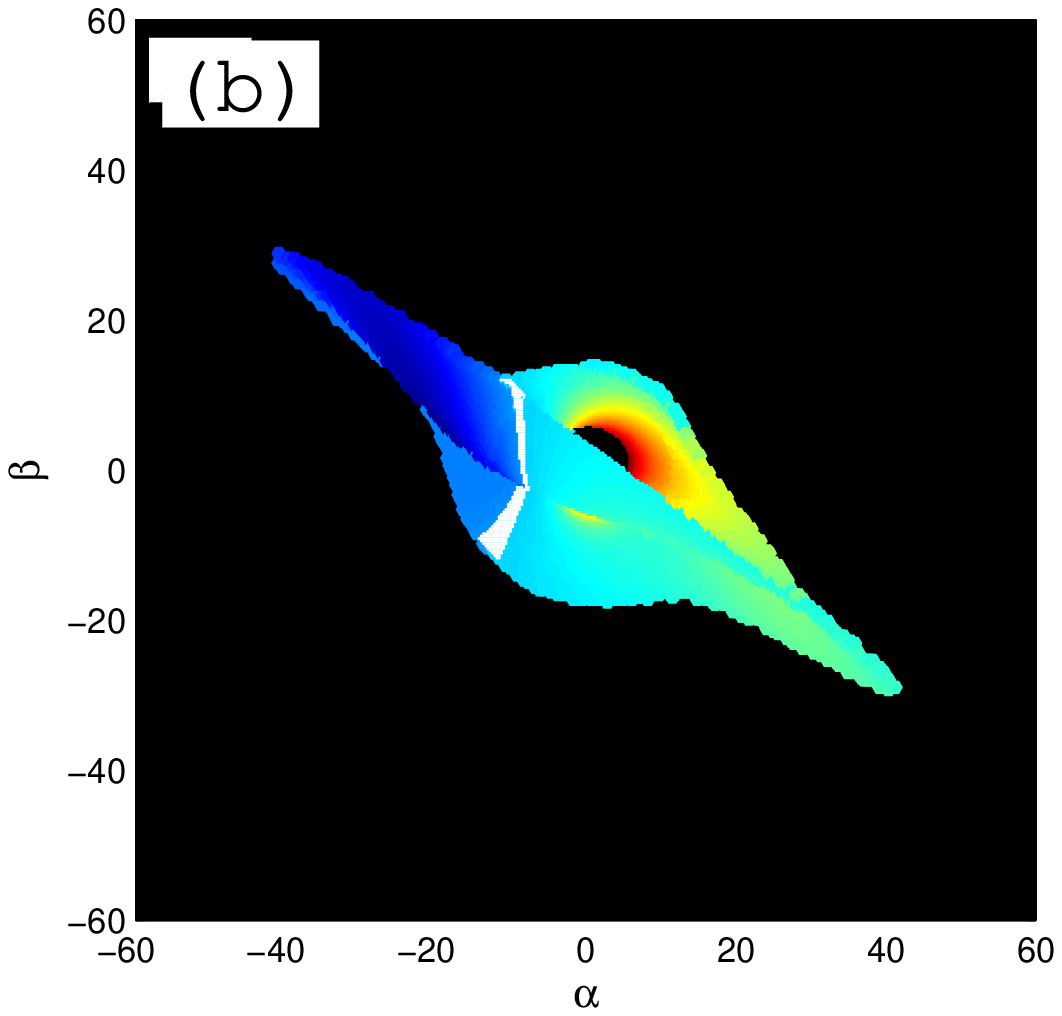,width=0.3\linewidth} &
\epsfig{file=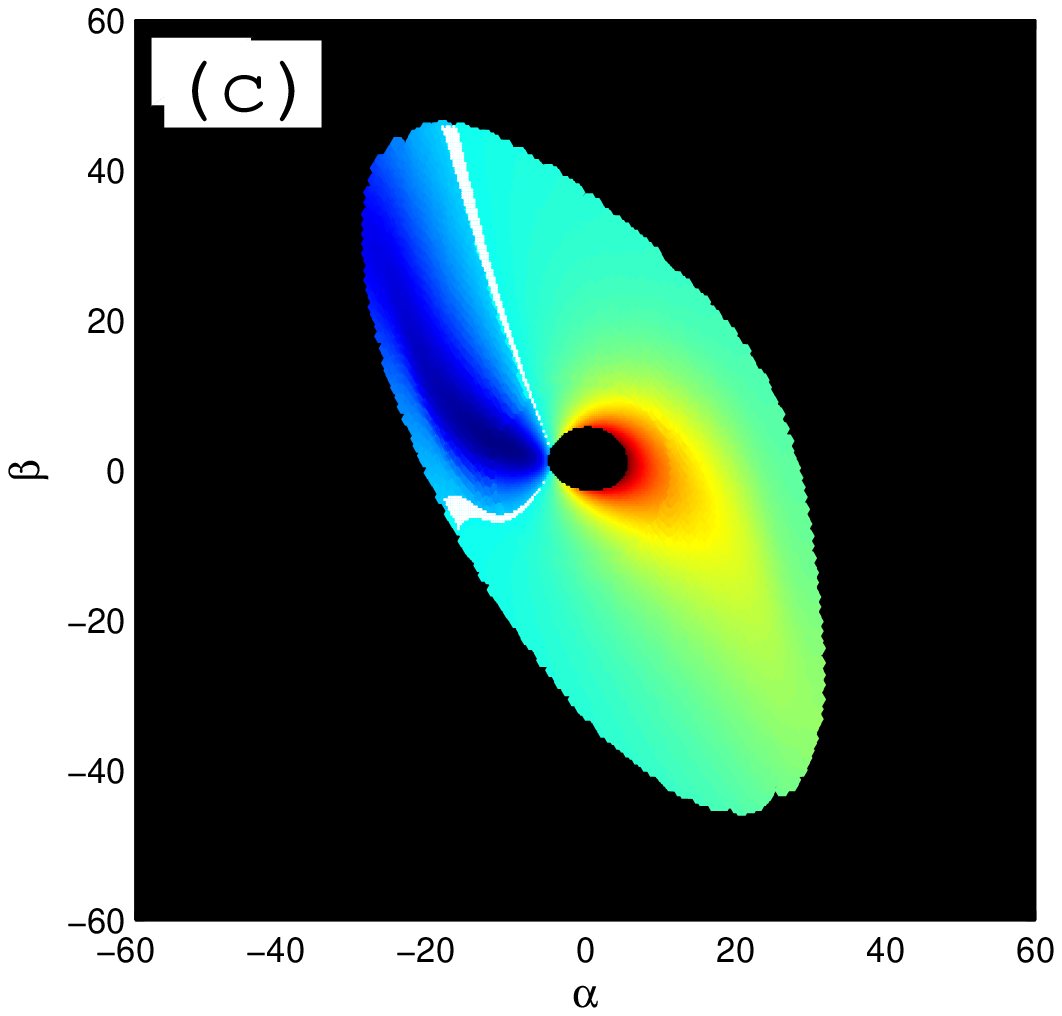,width=0.3\linewidth} \\
\epsfig{file=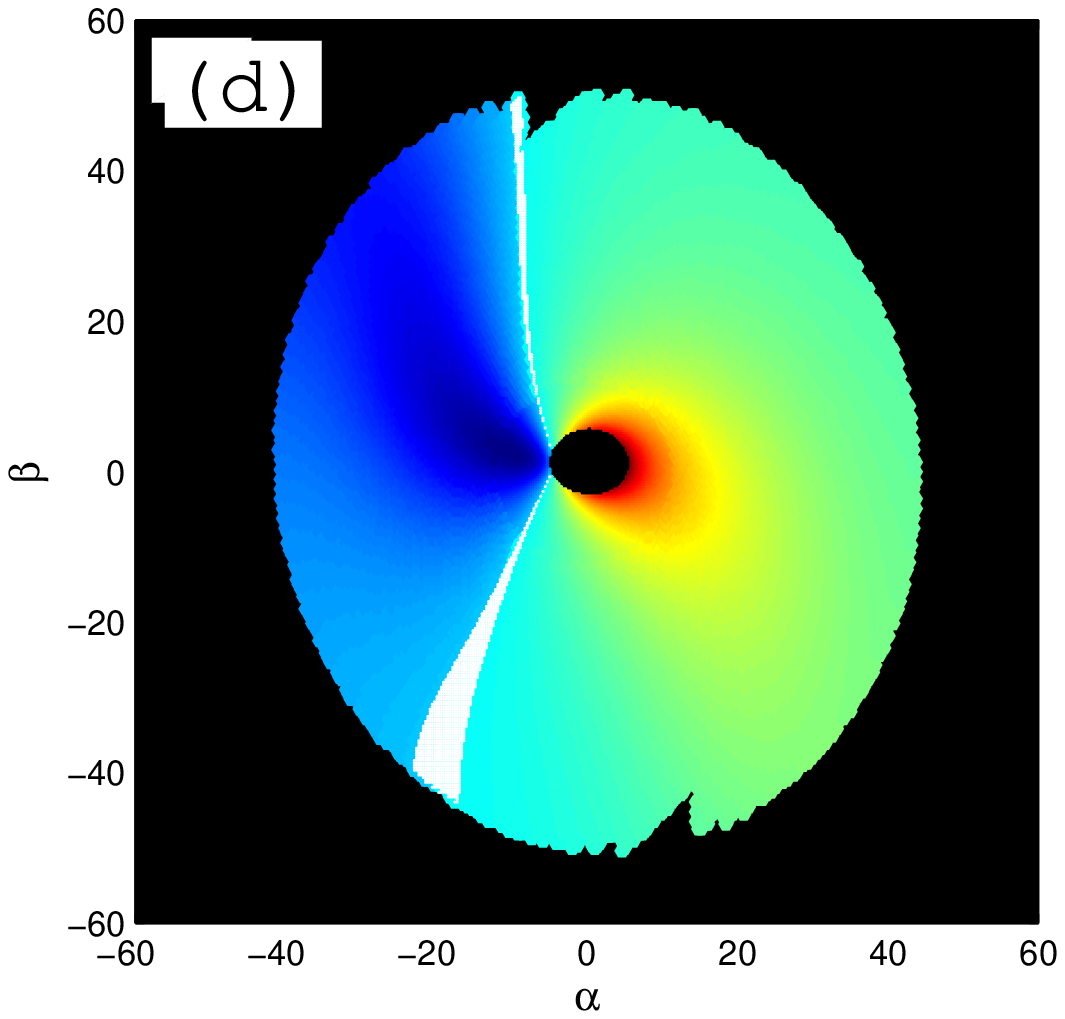,width=0.3\linewidth} &
\epsfig{file=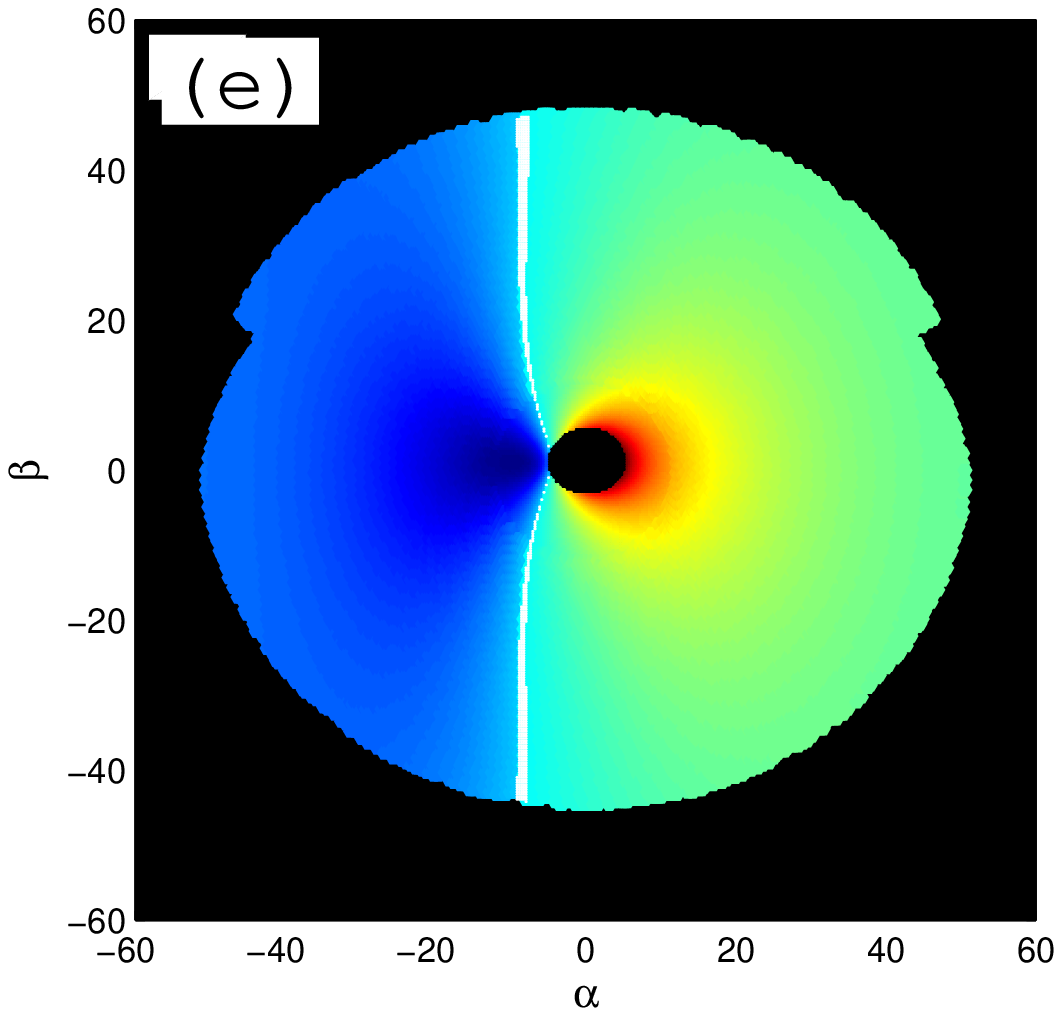,width=0.3\linewidth} &
\epsfig{file=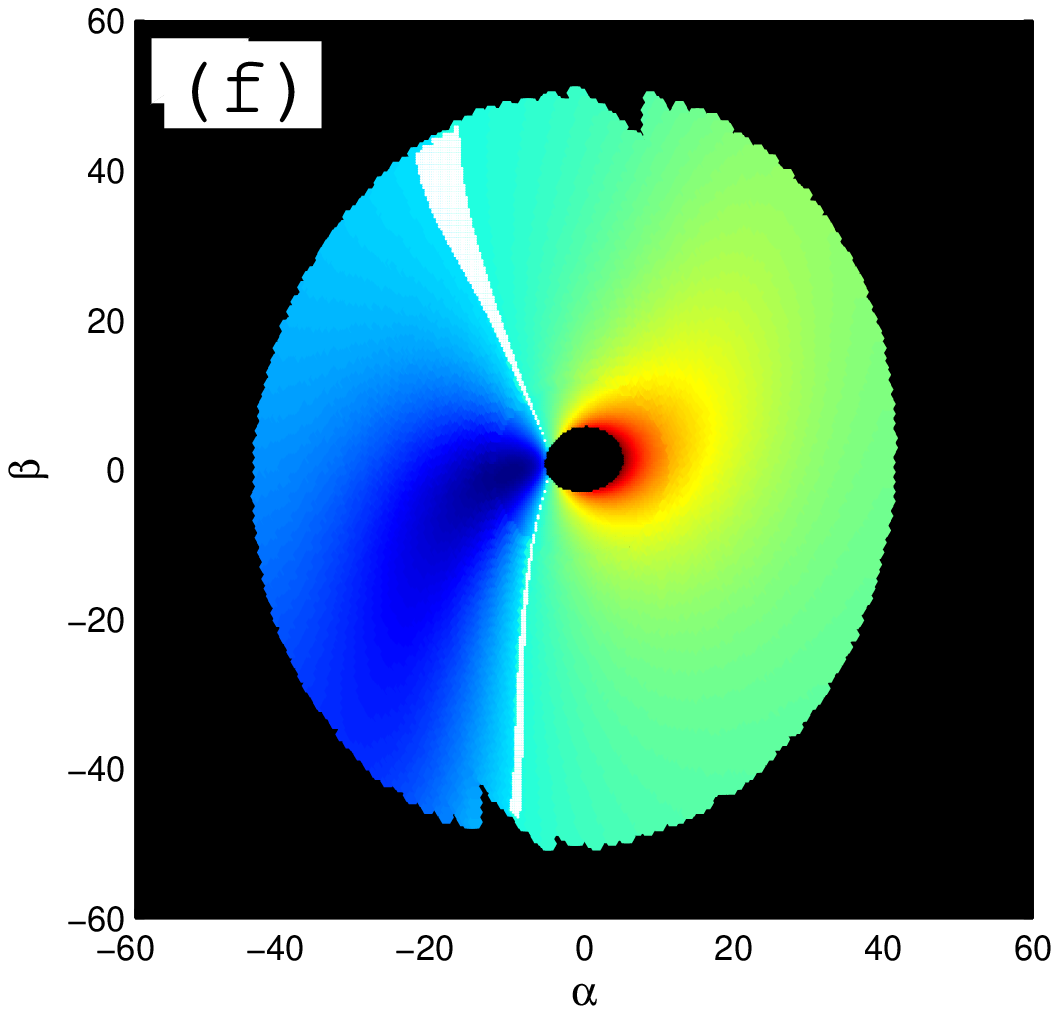,width=0.3\linewidth} \\
\epsfig{file=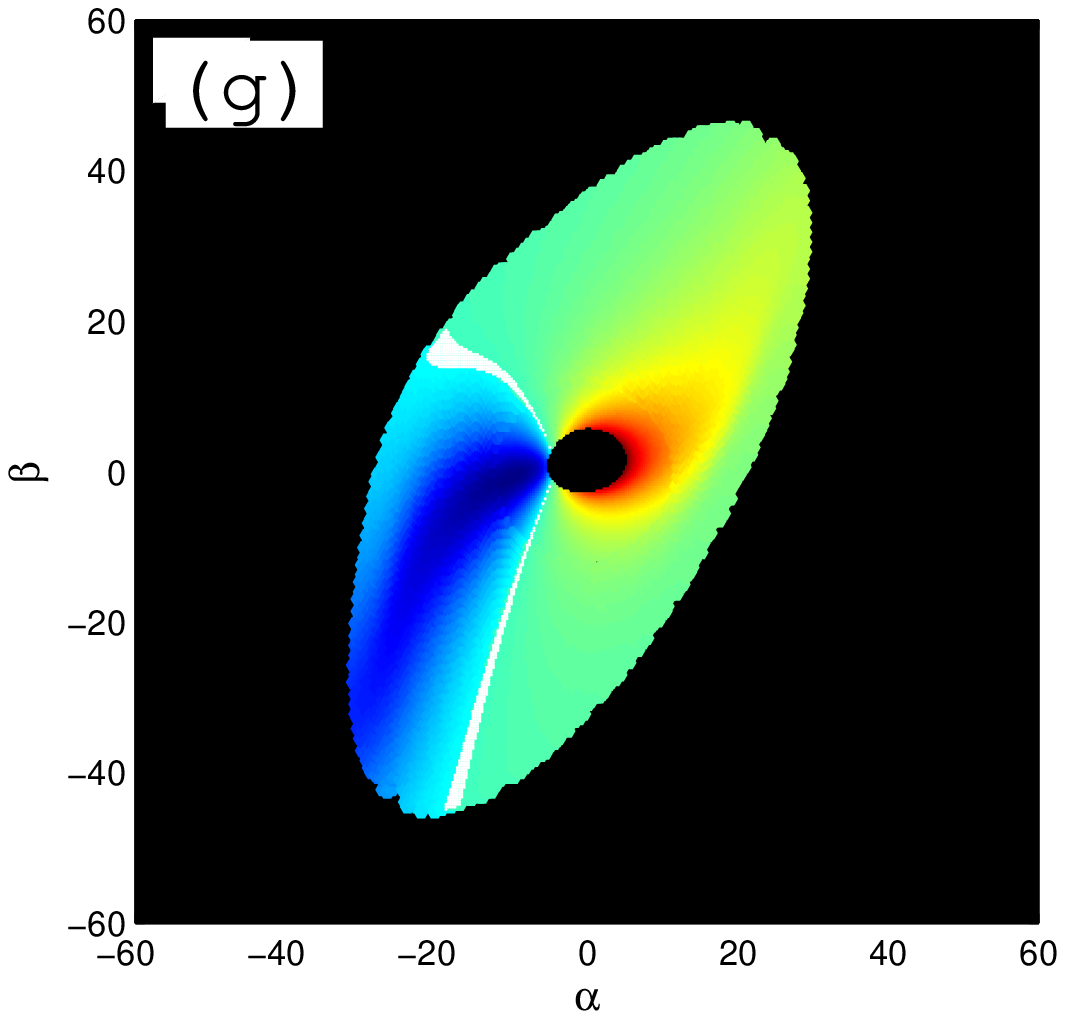,width=0.3\linewidth} &
\epsfig{file=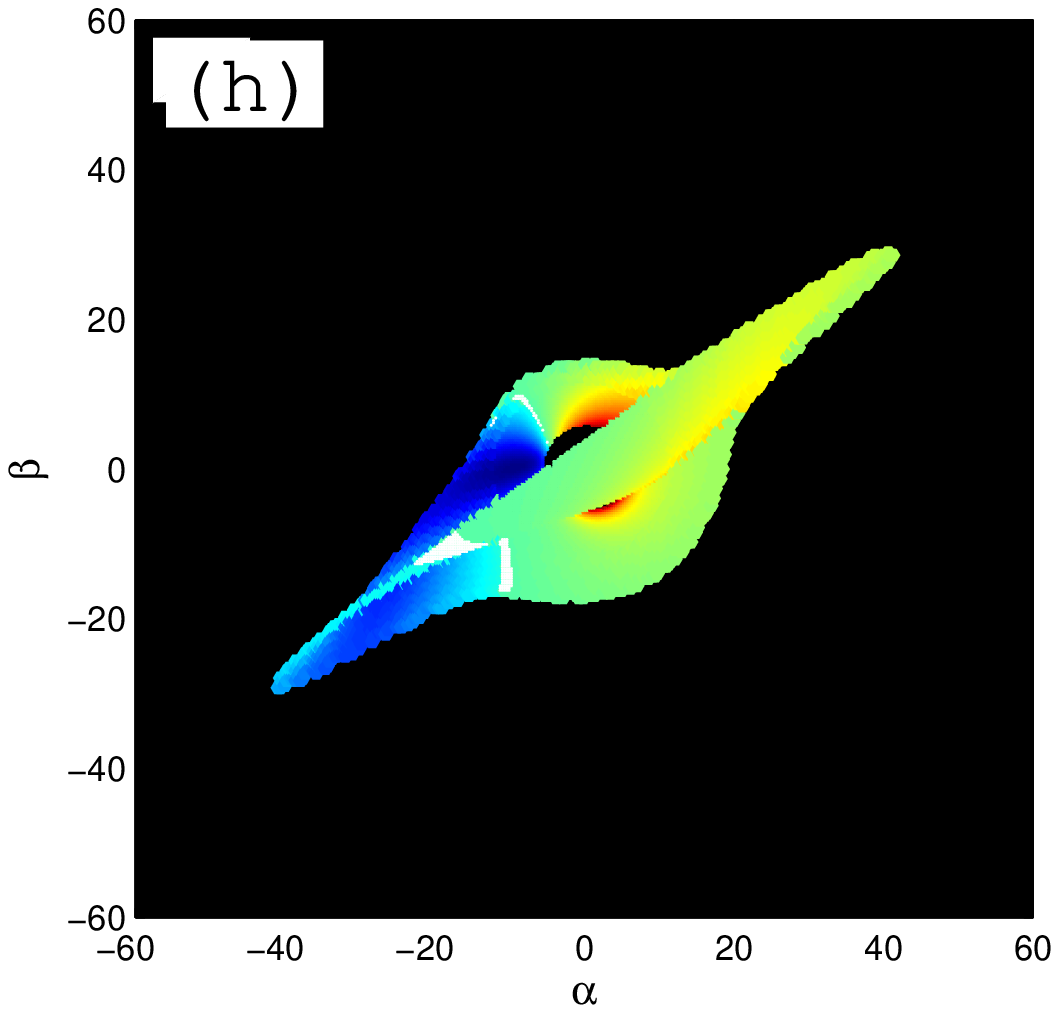,width=0.3\linewidth} &
\epsfig{file=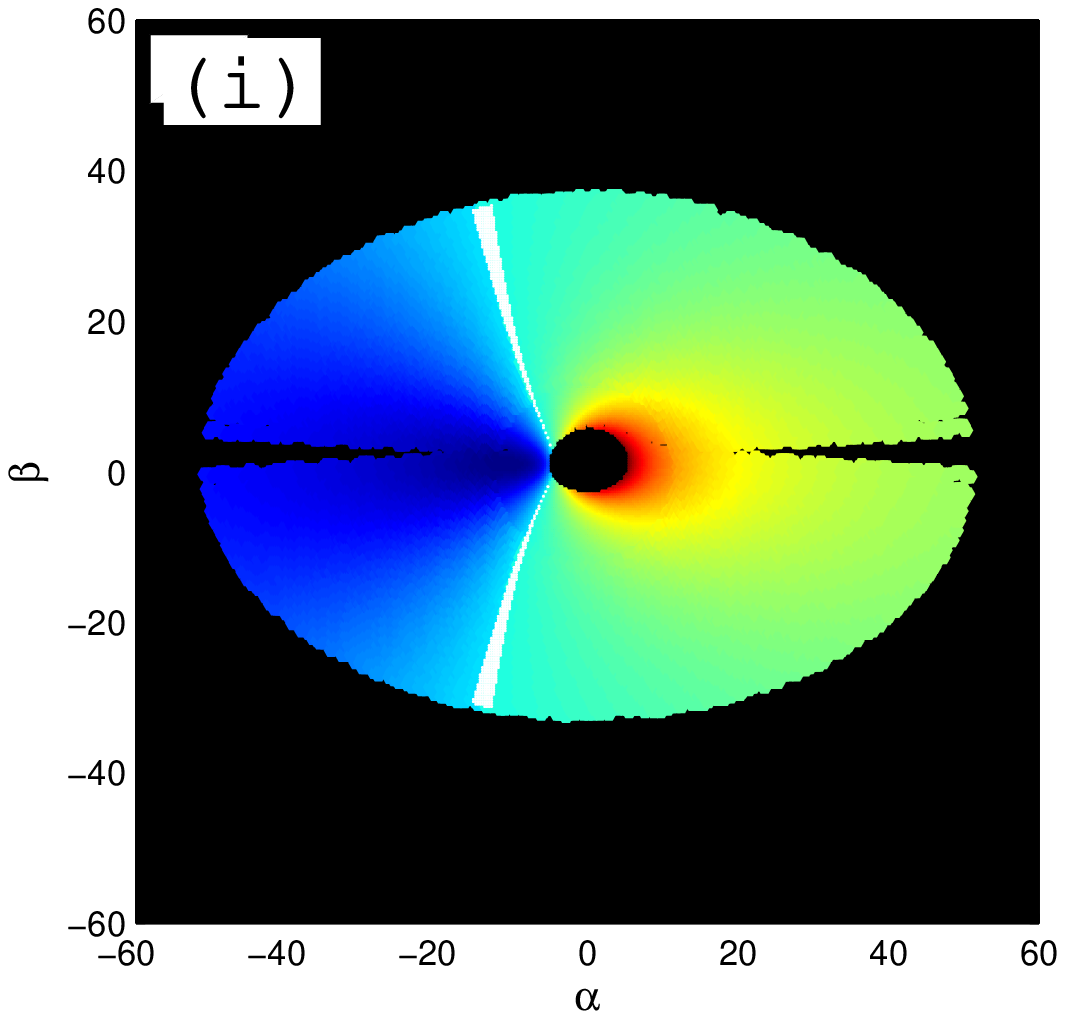,width=0.3\linewidth} \\
\end{tabular}
\end{center}
\caption{Images of a geometrically thin and optically thick warped disk around a
 Kerr black hole. The spin parameter $a$ is $0.5M$, and the observer's inclination angle
 $\theta$ is $50^{\circ}$ measured from the spin axis. The warping parameters are the same as
 previous figures ($n_1=0$, $n_2=1$, $n_3=0.95$). The inner and outer radius are $4.23~r_g$
 and $50~r_g$. $\alpha$ is the horizontal axis of the observer's photographic
 plate, and $\beta$ is the vertical axis. The false color contour maps show the ratio of the
 observed energy to the emitted energy. The blue shaded areas represent the regions of the
 disk where photons emitted are blue-shifted, while the red shaded areas represent the regions
 where photons are red-shifted. The large areas colored with cyan and green represent the regions
 where photon energies are weakly blue and red-shifted. The white areas are the zero-shift
 regions, where the gravitational red-shift is balanced with the Doppler blue-shift.
 Panels (a)-(h) contain the line profiles seen from different azimuthal angles $\phi$ of
 $0^{\circ}$, $45^{\circ}$, $90^{\circ}$, $135^{\circ}$, $180^{\circ}$, $225^{\circ}$, $270^{\circ}$ and
 $315^{\circ}$. Panel (a) shows the $\phi=0^{\circ}$ case which is defined as the observer viewing
 from the direction furthest from the lowest point of the disk. Panel (i) shows the disk image
 from a standard flat disk for comparison.}
\label{fig:k5IM50}
\end{figure}

 The line profile is sensitive to the inclination angle of the observer. For the low
 inclination angle $\theta=10^{\circ}$ (red), the line profile in panels (a)-(h) tends
 to be a single broad peak with the same endpoints of the red and blue tails, and
 almost all of the photons are red-shifted due to strong gravitational red-shift and
 weak Doppler blue-shift. Those line profiles are similar to the flat disk line profile
 in panel (i). With the increase of the inclination angle (blue and black), the two-peak or
 multi-peak structures start to emerge. Compared with the flat disk case where the two
 peaks are shown (the blue peak is always brighter than the red one), the line profiles
 from the warped disk present considerable changing of the features, for example the
 red peak can be comparable or stronger than the blue one in some cases. The deviations
 are produced by the warping structure and shadowing effect. which can only be seen
 from the images of the disk in Fig.~\ref{fig:k5IM50} (for $\theta=50^{\circ}$).

 For the large inclination angles ($\theta=50^{\circ}$ and $\theta=70^{\circ})$, the
 shape of the line profiles is also sensitive to the azimuthal angle. This is because the
 observable regions of the disk are changing with the precession of the disk over a
 long period (see images in Figs.~\ref{fig:k5IM50} for $\theta=50^{\circ}$). It appears that the
 line profiles are less sensitive to the low inclination angles ($\theta=10^{\circ}$).
 Since at the low inclination, the inner region of the disk is almost face-on (for the
 twist-free warped disk the innermost area coincides with the equatorial plane) and the
 shadowing effect due to the warping is less important. We note that the black lines in panel (a)
 ($\phi=0^{\circ}$) and panel (e) ($\phi=180^{\circ}$) for $\theta=70^{\circ}$ are similar,
 and the relatively narrower tails for $\phi=0^{\circ}$ is due to the fact that part of the
 strongly blue-shifted and red-shifted area is shadowed.

\begin{figure}[h]
\begin{center}
\includegraphics[width=1.0\textwidth]{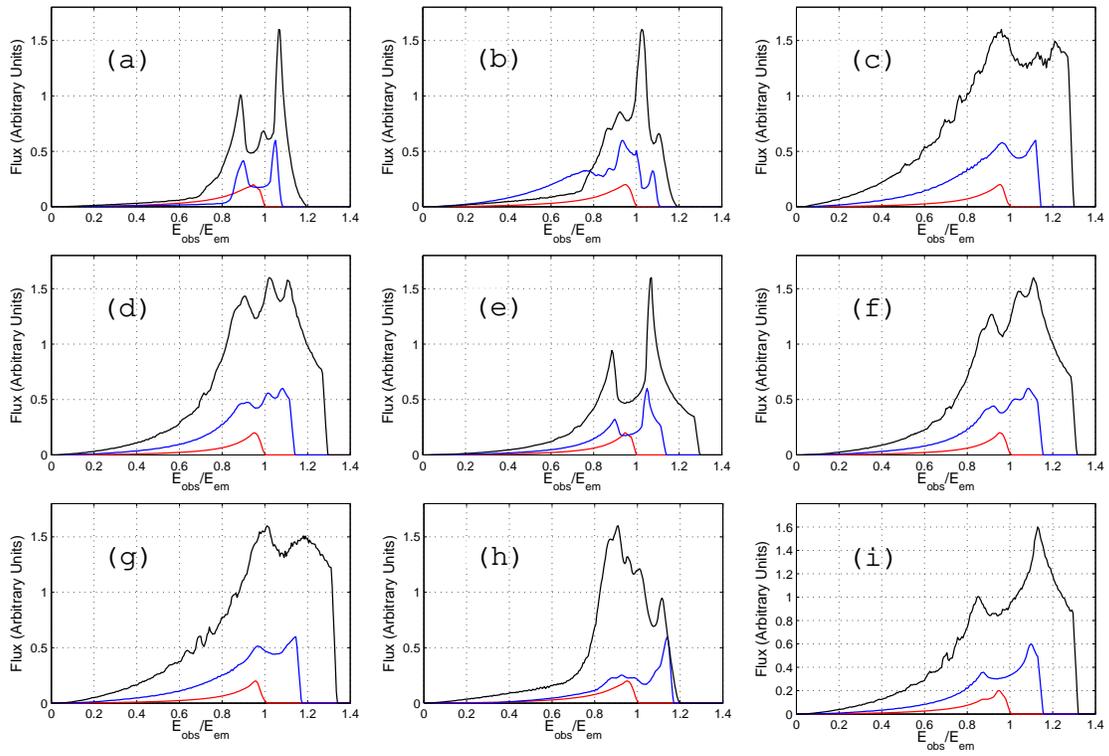}
 \caption{Line profiles as in Fig.~\ref{fig:dell_kerr5SF1357} with $a=0.998M$,
 $r_{in}=1.23~r_g$ and $r_{out}=50~r_g$.
 }
\label{fig:dell_kerr9SF1357}
\end{center}
\end{figure}

 Figure~\ref{fig:dell_kerr9SF1357} shows the line profiles from the accretion disk with
 the same warping parameters $n_1$, $n_2$, and $n_3$ as in Figure~\ref{fig:dell_kerr5SF1357}
 but for a more relativistic system, where the black hole has an extreme spin parameter
 $a=0.998M$ \citep{1974ApJ...191..507T}. The inner radius $r_{\text{in}}$ for the accretion disk is
 only $1.23~r_g$. The general features mentioned for Figure~\ref{fig:dell_kerr5SF1357} hold here.
 In addition, the effect of the gravitational red-shift and light bending is more significant.
 Some photon trajectories which are initially retrograde may be forced to
 turn back to be prograde by the frame dragging effect. In this more relativistic system,
 the red tails are rather long and extend to as far as $g=0.1$, while the blue tails end
 at about the same places as in Figure~\ref{fig:dell_kerr5SF1357}, although they are cut off
 more sharply for largest inclinations ($\theta=70^{\circ}$). However, the line profiles
 around $g=1$ (the rest energy) are very similar to Figure~\ref{fig:dell_kerr5SF1357},
 since the contribution to the flux here is mainly from the rings in the outer region of
 the disk ($r>10~r_g$), where the frame dragging effect is much smaller ($w\propto r^{-3}$).

\begin{figure}[h]
\begin{center}
\includegraphics[width=1.0\textwidth]{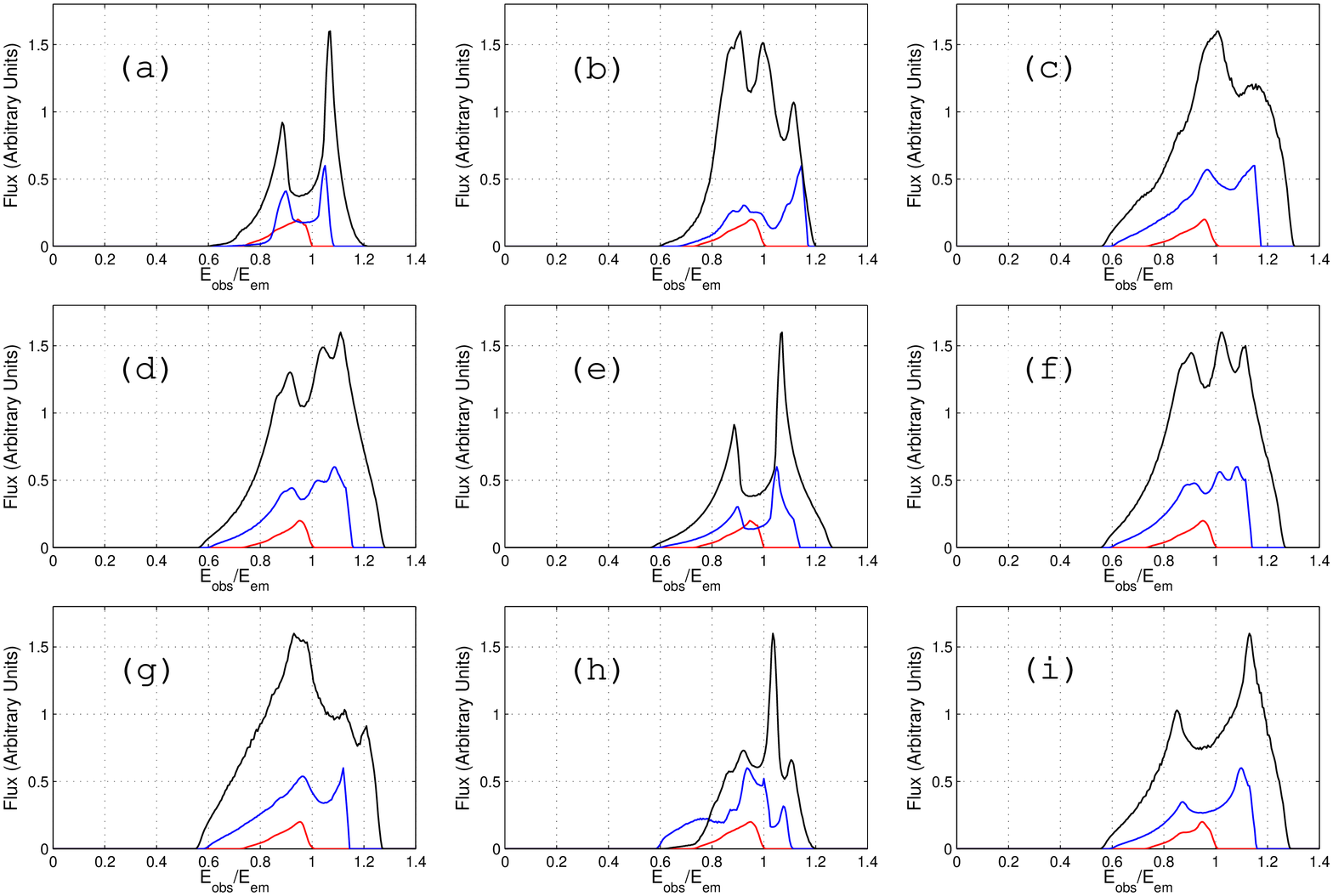}
\caption{Line profiles as in Fig.~\ref{fig:dell_kerr5SF1357} with $a=-0.5M$,
 $r_{in}=7.55~r_g$ and $r_{out}=50~r_g$.
 }
\label{fig:dell_kerr5mSF1357}
\end{center}
\end{figure}

 Figure~\ref{fig:dell_kerr5mSF1357} shows the line profiles from the accretion disk with
 the same warping and spin parameters as in 
 Figures~\ref{fig:dell_kerr5SF1357}, except we choose the disk to rotate around the black 
 hole in a retrograde orbit ($\epsilon=-1$). The retrograde orbiting of the disk
 will increase its inner radius $r_{\text{in}}$ to a larger value $7.55~r_g$, which reduces the
 contribution from the innermost part of the disk so that the red and blue tails become
 slimmer than in Fig.~\ref{fig:dell_kerr5SF1357}. In addition the patterns
 are anti-symmetric about $180^{\circ}$ compared to Fig.~\ref{fig:dell_kerr5SF1357}.

\subsubsection{Changing of the emissivity index $q$}

 The detailed irradiation law is unknown and the emissivity of the iron line in the inner
 region of the disk is poorly understood so far. For simplicity, we assume that the emissivity
 is a power-law function of radius with the emissivity index as the power-law index
 (i.e. $\varepsilon\sim(r/M)^{q}$).

\begin{figure}[h]
\begin{center}
\includegraphics[width=1.0\textwidth]{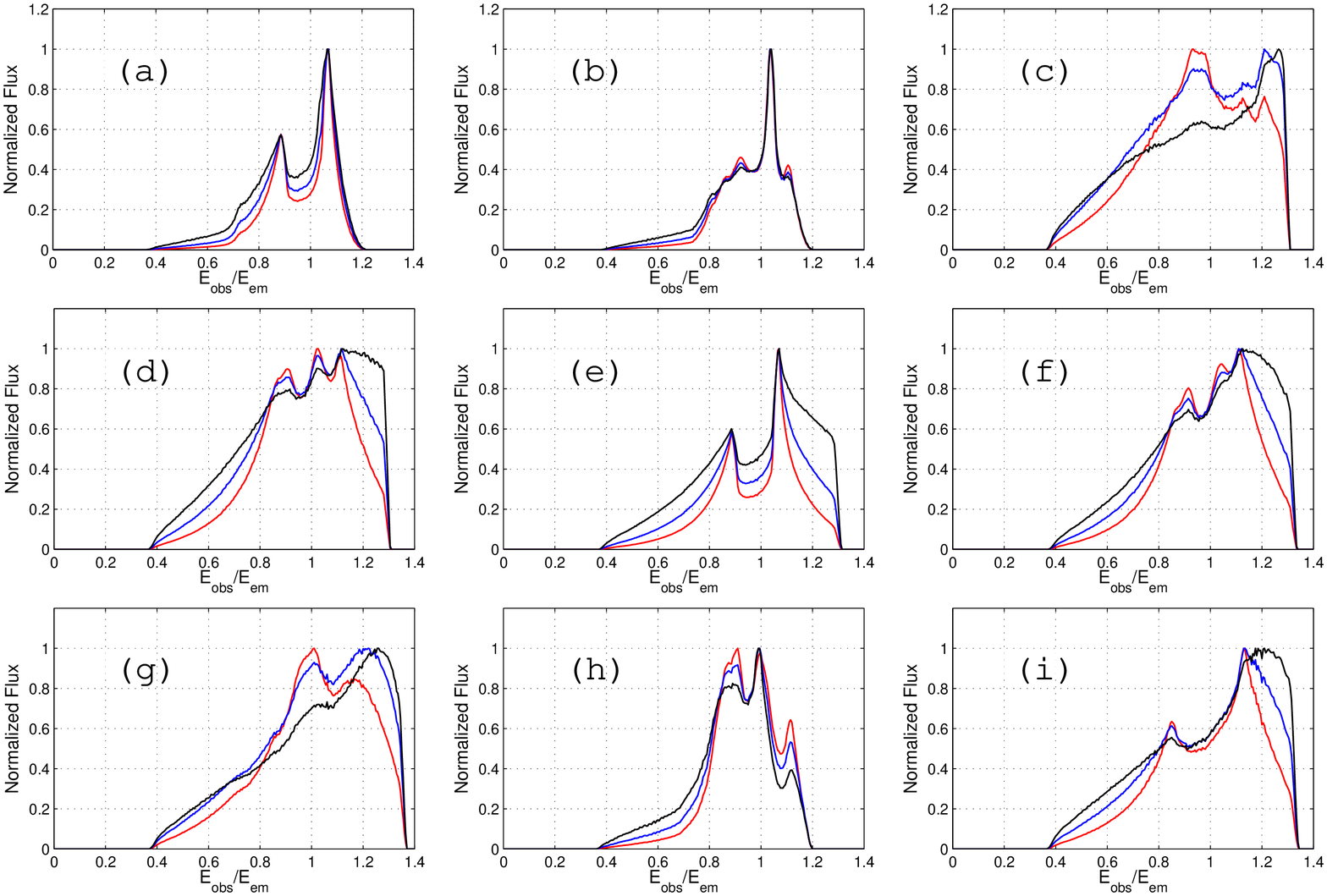}
\caption{Line profiles from a warped disk with $n_1=0$, $n_2=1$, $n_3=0.95$,
 $r_{in}=4.23~r_g$ and $r_{out}=50~r_g$. The spin parameter $a$ of the Kerr black hole is $0.5M$. The emissivity index
 $q$ is taken to be $-2$. The horizontal axis is the $g$ factor, i.e. the observed photon energy
 per unit iron K-shell photon energy; the vertical axis is flux in arbitrary units.
 The inclination angle $\theta$ is $70^{\circ}$ measured from the spin axis of
 the black hole. In each panel, three line profiles represent emissivity index
 $q$ of $-2$ (red), $-2.5$ (blue), and $-3$ (black)
 respectively. Panels (a)-(h) contain the line profiles seen from different azimuthal angles $\phi$ of
 $0^{\circ}$, $45^{\circ}$, $90^{\circ}$, $135^{\circ}$, $180^{\circ}$, $225^{\circ}$, $270^{\circ}$ and
 $315^{\circ}$. Panel (a) shows the $\phi=0^{\circ}$ case which is defined as the observer viewing
 from the direction furthest from the lowest point of the disk. Panel (i) shows the line profiles
 from a standard flat disk for comparison.
 }
\label{fig:dell_kerr5SFabc}
\end{center}
\end{figure}

 Figure~\ref{fig:dell_kerr5SFabc} shows the line profiles from the warped disk with the
 same warping parameters and the spin parameter as in Figure~\ref{fig:dell_kerr5SF1357}.
 The difference is that the inclination angle $\theta$ in all cases here are chosen as $70^{\circ}$
 measured from the black hole spin axis. In each panel, the three colored lines represent the
 line profiles with the emissivity index $q$ being $-2$ (red), $-2.5$ (blue), and $-3$ (black)
 respectively. Panels (a)-(h) contain the line profiles seen from different azimuthal
 viewing angles as in Figs.~\ref{fig:dell_kerr5SF1357}--\ref{fig:dell_kerr5mSF1357}, and
 panel (i) shows the line profiles from the standard flat disk for comparison.
 All the line profiles are normalized to unity.

 Generally speaking, smaller emissivity index indicates relatively larger contribution
 from the inner region of the disk, if we do not consider the shadowing effect which
 prevents part of the inner region to be seen in some cases. The inner regions contribute
 mostly to the flux at blue and red tails while the outer region to that around $g=1$ (the
 rest energy). Therefore, for smaller emissivity index (black than red), the flux in the
 red/blue tails is more intensive than that for larger emissivity index. However
 the shadowing effect is pronounced in some cases. For example in panel (a) ($\phi=0^{\circ}$),
 part of the most blue-shifted region is unobservable due to the shadowing, the blue
 tails for three different $q$ are almost identical. Moreover, they are less blue-shifted
 ($g\simeq 1.2$) and are not steep as others. Similar feature appears in
 panel (b) ($\phi=45^{\circ}$) and panel (h) ($\phi=315^{\circ}$) too. In panel (e)
 ($\phi=180^{\circ}$), the disk is fully observed (least shadowing), the main feature of the
 line profile is similar to the flat disk where there is no shadowing at all.
 In panels (c)--(g), where the shadowing effect is less important, the blue tails for
 smaller emissivity index are steeper, and it is a common feature shared by both the warped disk
 and the flat disk. Since the nature of the central black hole, the dynamic structure of
 the disk and the orientation to the observer are the same for profiles in each panel,
 the blue and red tails representing different emissivity index end at the same places.

\begin{figure}[h]
\begin{center}
\includegraphics[width=1.0\textwidth]{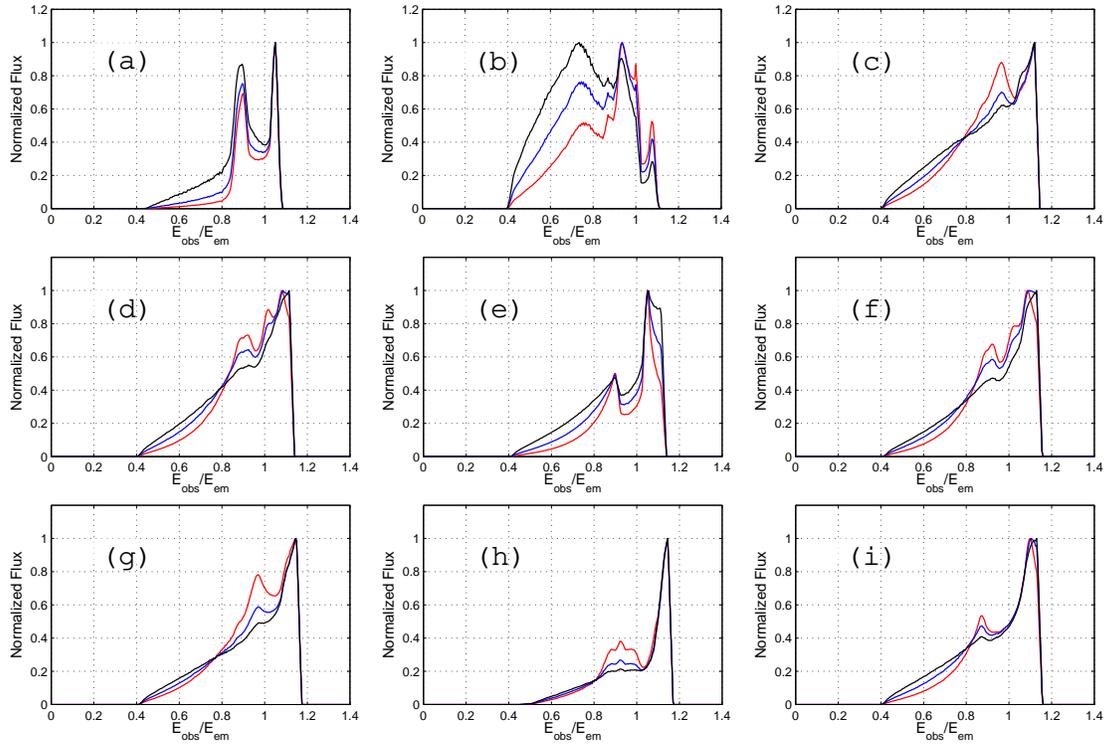}
\caption{Line profiles as in Fig.~\ref{fig:dell_kerr5SFabc} with
 $\theta=50^{\circ}$, $r_{in}=4.23~r_g$ and $r_{out}=50~r_g$.
 }
\label{fig:dell_kerr5SF50abc}
\end{center}
\end{figure}

 Figure~\ref{fig:dell_kerr5SF50abc} shows the line profiles from the warped disk with the
 same warping parameter and the spin parameter as in Figure~\ref{fig:dell_kerr5SFabc}, but the
 inclination angle $\theta$ here are chosen as $50^{\circ}$. The features mentioned
 for Figure~\ref{fig:dell_kerr5SFabc} remain true here. Furthermore the blue tails are
 reduced to smaller values ($g<1.2$) and cut off more steeply due to smaller Doppler blue-shift for
 this smaller inclination angle; while the red tails end at about the same value ($g\simeq 0.4$),
 because they are mainly from the gravitational red-shift which is not sensitive to the inclination
 angle.

\begin{figure}[h]
\begin{center}
\includegraphics[width=1.0\textwidth]{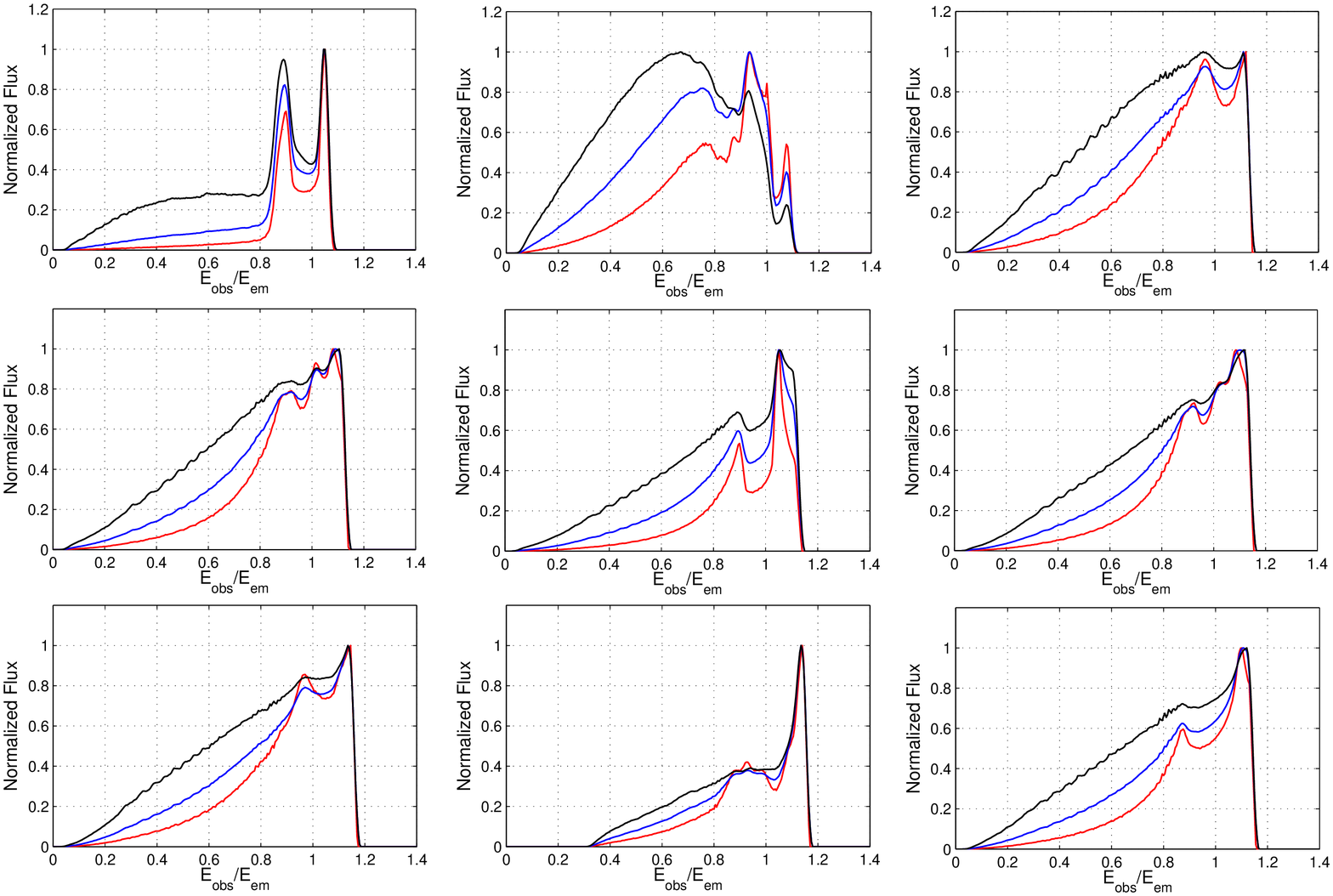}
\caption{Line profiles as in Fig.~\ref{fig:dell_kerr5SF50abc} with
 $a=0.998$, $\theta=50^{\circ}$, $r_{in}=1.23~r_g$ and $r_{out}=50~r_g$.
 }
\label{fig:dell_kerr9SFabc}
\end{center}
\end{figure}

 Figure~\ref{fig:dell_kerr9SFabc} shows the line profiles from the warped disk with the same
 warping parameter and inclination angle as in Figure~\ref{fig:dell_kerr5SF50abc} but with the
 spin parameter $a=0.998M$. For such extreme spin rate, the inner radius of the accretion disk
 can be very close to the event horizon ($r_h=1.0632~r_g$), where the gravitational red-shift
 dominates. Therefore the red tails in Figure~\ref{fig:dell_kerr9SFabc} are much longer than
 those in Figure~\ref{fig:dell_kerr5SF50abc}. For the reason we motioned above, smaller emissivity
 index usually produces more intensive red tails, the red tails for $q=-3$ (black) is stronger
 than those of the other two lines. In extreme case, for example panel (a), a bump appears
 in the red tail of $q=-3$.

\section{Discussion and Conclusions}\label{sec:conc}

 To date there are two practical methods for measuring the spin of astrophysical black holes,
 namely the X-ray continuum fitting and the relativistic iron lines. The continuum-fitting
 method so far can only be applied to stellar-mass black holes \citep{2011arXiv1101.0811M},
 whereas the relativistic iron lines can be used as a probe to both the spins of stellar-mass
 black holes in X-ray binaries and supermassive black holes in active galactic nuclei. Furthermore
 if the accretion disks in these systems are warped, it is possible that the relativistic iron
 line profiles may serve as a way to diagnose the structures of the disks.

 In this paper we have developed a method for calculating the emission line profiles from the
 warped accretion disks orbiting around rotating black holes. It essentially generalizes previous
 calculations by including the black hole spin, the disk warping and the shadowing effect at the
 same time. We have also presented the disk images which can help us
 understand the general features of the line profiles from different disk and black hole
 systems. The detailed calculation concerns the computation of the trajectories of the photons
 emitted from the disk and the velocity field of the plasma elements orbiting around the center
 black hole. The formalism described above makes no assumption about the emitting energy of
 the photon. Therefore it can be applied to any emission lines which could be detected in the
 immediate vicinity of the black hole, such as the carbon, nitrogen, and oxygen lines although
 they are not as prominent as the iron lines.

 We assumed a set of warping parameters for the geometry of the twist-free warping disk and
 Keplerian orbit for the motion of the plasma elements as examples. Even for this simplified
 system we have found a number of interesting phenomena:
 (1) Different from the two-peak feature generally found in flat disks, the line profiles from
 the warped disk show multiple peaks.
 (2) The line profiles for some cases present sharper red tails and/or softer blue
 tails, which may provide an alternative explanation to observations for some Seyfert 1 galaxies
 \citep{1997ApJ...477..602N} and Seyfert 2 galaxies \citep{1997ApJ...488..164T}.
 (3) When the disks are orbiting around a highly spinning black hole, a rather long red tail is
 a common feature shared between flat disks and warped disks, if the shadowing effect is not important.
 (4) At low inclinations, the line profiles from different azimuthal angles differ very
 little from each other, indicating that the warping has little influence on the line profile
 and it may not provide much information for diagnosing the disk structure.
 (5) The line profile is also sensitive to the illumination law: smaller emissivity index $q$
 generally indicates stronger red and blue tails. In some extreme cases, red bumps may grow
 in the red tails.
 (6) It may not be true that a sharp blue tail is a common feature in all axisymmetric disk
 models, and is expected for time-integrated profiles even if the disk has strong asymmetry
 or inhomogeneities \citep{1998Natur.391...54B}. The shadowing we encounter here may cause
 a soft blue tail in the integrated profile.
 (7) With the precession of the warped disk, time variations of line profiles may be present,
 and may be a possible observational signature for the warped disk.
 (8) Photometric variations of the line fluxes may be due to the varying amount of area of
 the disk facing to the observer.
 
 It is nontrivial to constrain the model parameters by fitting the theoretical line 
 profiles to observations. However, from the above results and discussions we can 
 see that generally the red tails of the line profiles provide the information
 about the spin of the black hole, while the blue tails the inclination angle. And the warping would 
 cause complicated multiple peaks (which depends on the warping parameters), and time-variation 
 of line profiles with the precession of disk. In the large inclination cases, the warping would 
 cause significant disk shadowing which may prevent all/part of the inner area to be seen.
 Smaller emissivity index $q$ normally indicates more intensive red and blue tails in the line
 profiles. 
 Our present work is an initiative of a systematic investigation about the impact of the disk warping 
 and shadowing on the line profile. It can be considered as an extension of previous work on line 
 profiles for warped disk \citep{2000MNRAS.317..880H,2008MNRAS.389..213W}. Our line profiles 
 distinguish from the line profiles from non-rotating black hole systems by presenting 
 longer red tails which could extend to $g=0.1$ for highly spinning cases. And the overall line
 feature is smooth and has similarity to the flat disk if the shadowing is not significant. These are   
 different from \citet{2000MNRAS.317..880H}, in which there are abrupt spike-like structures in some cases.
 We provide an alternative explanation to the variation of line profile which is 
 complementary to the picture of \citet{2001PASJ...53..189K}, who investigate the non-axisymmetric 
 patterns in the disk surface. However the non-axisymmetric patterns are unlikely to produce multiple 
 peak structure. Due to the axisymmetry, both the thick disk \citep{2007MNRAS.378..841W} and accretion 
 torus \citep{2007A&A...474...55F} would not show the time-variant of line profile with the 
 precession and the line profiles are usually single-peaked and double-peaked.

 Long-period photometric variation of X-ray is observed for many accreting systems. The precession
 of warped disk may be an explanation for this phenomenon. The time-scale characterizing the
 precession induced by tidal force of the companion star in XRB \citep{1999MNRAS.308..207W} is
 \begin{equation}
 t_{\Omega_p}=\frac{2\pi}{\Omega_p}=6.7\left(\frac{P_{orb}}{1d}\right)^2\left(\frac{1+q}{q}\right)
 \left(\frac{M}{1 M_{\odot}}\right)^{1/2}\left(\frac{R}{10^{11}\text{cm}}\right)^{-3/2}\text{d}  ~\,,
 \end{equation}
 where $P_{orb}$ is the orbital period of the binary, and $q$ is the ratio of the mass of the
 companion star to the central accreting object. The time-scale characterizing the precession
 induced by radiation can be found in \citet{1996ApJ...472..582M}. The typical precession period
 for an AGN may be much longer than that of the XRB. The purpose of showing this formula is to point out
 that apart from the time-averaged line profile (with typical explosion time longer than $10^6$\,s),
 it is possible to study the time variation
 of the iron line profiles on the precession timescale with the next generation space X-ray
 observatories which have sufficiently large collecting area.

 With the advance of the millimeter/sub-millimeter very-long baseline interferometer (VLBI), the
 proposed Earth-based \textit{Event Horizon Telescope} (EHT) \citep{2009astro2010S..68D} and the
 on-going space-VLBI (such as VSOP\footnote{http://www.vsop.isas.ac.jp/} and
 RadioAstron\footnote{http://www.asc.rssi.ru/radioastron/}) promise to provide microarcsecond
 imaging resolution, which is sufficient to resolve the event horizons and the inner
 regions of the accretion disks around a handful of supermassive black holes. The calculated disk
 images may help understand the possible asymmetric intensity profile, blue/red-shift,
 and other observational signatures caused by the warping of the accretion disk and the rotating
 of the black hole. In the following work we will try to constrain the model parameters by 
 fitting the theoretical profiles of the predicted spectra to actual data.

 Although we have given results for only one set of warping parameters, our formalism can be
 easily extended to more general disks. Work is underway on investigating systems with different
 warping parameters. By using the method of moments, we can explore the large warping parameter
 space ($n_1$, $n_2$, $n_3$) and study the distribution of the warping parameters in the moment
 space. In practice, it may be helpful to compare the positions of the prominent line peaks 
 in the warped disks with the corresponding peaks in the flat disks. And it would be easier than 
 fitting the overall line profiles \citep{2009ApJ...701..635M, 2011MNRAS.tmp.1522S}.
 Through this approaches we may connect the calculated line profiles with the current
 and/or future observations.

\section{Acknowledgment} We are grateful to an anonymous referee for constructive comments.
We also thank Xinlian Luo, Teviet Creighton and Frederick Jenet for helpful discussions and comments,
and Richard Price for directing us to several related publications and reviewing the manuscript.
This work was supported by the Natural Science Foundation of China (under grant number 10873008), 
and the National Basic Research Program of China (973 Program 2009CB824800).

\appendix

\section{Expression of $g$ for the warped disks}\label{sec:appA}


 This section describes how to derive the expression of $g$, the ratio of observed
 photon energy to the emitted energy for the warped disk.

\subsection{4-velocity of the particle}

 In each concentric ring, we let the angular position of the each plasma element be
 $\varphi=\Omega t$, with $t$ the coordinate time in the Kerr metric and the plasma
 particle's velocity components are
\begin{equation}
\mathbf{v'}=dx_{i'}/dt=v\left(-\sin\varphi,\cos\varphi,0\right) \,,
\end{equation}
 where $v=r\Omega=r(a+\epsilon\sqrt{r^3/M})^{-1}$, $\epsilon=1$ is for co-rotating/prograde
 orbiting, $\epsilon=-1$ is for counter-rotating/retrograde orbiting. In addition, we need
 to transfer the velocity into the coordinate frame in which the $z$ axis is aligned with
 the black hole spin. In Figure~\ref{fig:pic2}, the inclined $x'y'$ plane is the orbital
 plane, and the $y'$ axis coincides with the line of nodes and points at the ascending node.
 The $xy$ plane is the equatorial plane which has a tile angle $\beta$ relative to the $x'y'$
 plane. The projection of the $z'$ axis on the $xy$ plane has a twist angle $\gamma$ relative
 to the $x$ axis. We find the components $v_{i}$ in the $xyz$ coordinate basis using
 $v_{i}=T_{ij'}v_{j'}$, where
\begin{equation}\label{Tij}
T_{ij'}=\left( \begin{array}{ccc}
\cos\gamma\cos\beta & -\sin\gamma & -\cos\gamma\sin\beta \\
\sin\gamma\cos\beta & \cos\gamma & -\sin\lambda\sin\beta \\
\sin\beta & 0 & \cos\beta
\end{array} \right)\,,
\end{equation}
 from which we get
\begin{equation}\label{eq:nprimes}
\begin{split}
v_{x}&=v\left(-\cos\gamma\cos\beta\sin\varphi-\sin\gamma\cos\varphi\right) \\
v_{y}&=v\left(-\sin\gamma\cos\beta\sin\varphi+\cos\gamma\cos\varphi\right) \\
v_{z}&=v\left(-\sin\beta\sin\varphi\right) \,.
\end{split}
\end{equation}
 We then calculate the velocity components in the spherical coordinate by projecting
 the velocity $\mathbf{v}$ onto the spherical basis $\mathbf{e}_r$, $\mathbf{e}_{\theta}$,
 $\mathbf{e}_{\phi}$
\begin{equation}\label{eq:sphere}
\begin{split}
v_{r}&=0 \\
v_{\theta}&=v\left(-\cos\psi\cos\beta\sin\varphi\cos\theta+\sin\psi\cos\varphi\cos\theta+
\sin\beta\sin\varphi\sin\theta \right) \\
v_{\phi}&=v\left(\sin\psi\cos\beta\sin\varphi+\cos\psi\cos\varphi \right) \,.
\end{split}
\end{equation}
 where $\psi=\phi-\gamma$ is the azimuthal angle of the plasma element measured from
 the projection of the $x'$ axis on the $xy$ plane.

 In the Kerr spacetime, if we set $\tau$ as the proper time of the plasma element, it satisfies
 \begin{equation}\label{eq:particle}
 -d\tau ^2=ds^2=-\Sigma\Delta A^{-1}dt^2+\sin^2\theta A\Sigma^{-1}
 \left(d\phi-\omega dt\right)^2+\Sigma\Delta^{-1} dr^2+\Sigma d\theta^2\,,
 \end{equation}
 and we divide both sides by $dt^2$
\begin{equation}\label{eq:dtaudt2}
-\frac{d\tau^2}{dt^2}=-\Sigma\Delta A^{-1}+\sin^2\theta A\Sigma^{-1}
\left(\dot{\phi}-\omega\right)^2+\Sigma\dot{\theta}^2  \,.
\end{equation}
 Therefore we can write the 4-velocity of the plasma element as follow,
\begin{equation}\label{eq:4velocity}
\begin{split}
u^{0}&=\frac{dt}{d\tau}=\left(\Sigma\Delta A^{-1}-\sin^2\theta A\Sigma^{-1}\left(\dot{\phi}
-\omega\right)^2-\Sigma\dot{\theta}^2\right)^{-1/2}  \\
u^{r}&=\frac{dr}{d\tau}=\frac{dr}{dt}\frac{dt}{d\tau}=\dot{r}u^{0}=0  \\
u^{\theta}&=\frac{d\theta}{d\tau}=\frac{d\theta}{dt}\frac{dt}{d\tau}=\dot{\theta}u^{0}=u^{0}\frac{v_{\theta}}{r}  \\
u^{\phi}&=\frac{d\phi}{d\tau}=\frac{d\phi}{dt}\frac{dt}{d\tau}=\dot{\phi}u^{0}=u^{0}\frac{v_{\phi}}{r\sin\theta} \,.
\end{split}
\end{equation}
 which are identical to Eq.~\ref{eq:u0f} in Sec.~\ref{sec:struc2} if insert
 Eq.~\ref{eq:sphere} into Eq.~\ref{eq:4velocity}. We note that the radial
 component of the velocity is zero because of we have assumed that the particle
 follows Keplerian orbit. And the term under the square root of Eq.~\ref{eq:4velocity}
 is always positive if $r>r_{\text{ISCO}}$.

\subsection{expression for $g$}

 The 4-velocity of the particle are expressed by the Boyer-Lindquist coordinates
 in Eq.~\ref{eq:4velocity}, we then derive the expression for $g$. As defined
\begin{equation}
g=\frac{E_{\text{obs}}}{E_{\text{em}}}=\frac{(u^{\mu}p_{\mu})_{\text{obs}}}{(u^{\mu}p_{\mu})_{\text{em}}}=(1+z)^{-1} ~\,.
\end{equation}
 The observer rest at infinity has 4-velocity $p_{\mu}$=(-1,\,0,\,0,\,0), thus in
 the numerator $(u^{\mu}p_{\mu})_{\text{obs}}=E$; while in the denominator
\begin{equation}
(u^{\mu}p_{\mu})_{\text{em}}=-Eu^0\pm Eu^0\sqrt{\Theta}v_{\theta}/r+L_zu^0v_{\phi}/\sin\theta ~\,,
\end{equation}
 therefore we get
\begin{equation}
g=\frac{1}{u^0(-1\pm\sqrt{\Theta}v_{\theta}/r+\xi v_{\phi}/\sin\theta)} ~\,,
\end{equation}
 where the sign is determined by the zenithal emitting direction.

\bibliographystyle{apj}
\bibliography{emission_fin3}

\begin{thebibliography}{60}
\expandafter\ifx\csname natexlab\endcsname\relax\def\natexlab#1{#1}\fi

\bibitem[{{Abramowicz} {et~al.}(2010){Abramowicz}, {Jaroszy{\'n}ski}, {Kato},
  {Lasota}, {R{\'o}{\.z}a{\'n}ska}, \& {S{\c a}dowski}}]{2010A&A...521A..15A}
{Abramowicz}, M.~A., {Jaroszy{\'n}ski}, M., {Kato}, S., {Lasota}, J.,
  {R{\'o}{\.z}a{\'n}ska}, A., \& {S{\c a}dowski}, A. 2010, \aap, 521, A15+

\bibitem[{{Bardeen} \& {Petterson}(1975)}]{1975ApJ...195L..65B}
{Bardeen}, J.~M., \& {Petterson}, J.~A. 1975, \apjl, 195, L65+

\bibitem[{{Barr} {et~al.}(1985){Barr}, {White}, \&
  {Page}}]{1985MNRAS.216P..65B}
{Barr}, P., {White}, N.~E., \& {Page}, C.~G. 1985, \mnras, 216, 65P

\bibitem[{{Brenneman} \& {Reynolds}(2006)}]{2006ApJ...652.1028B}
{Brenneman}, L.~W., \& {Reynolds}, C.~S. 2006, \apj, 652, 1028

\bibitem[{{Bromley} {et~al.}(1997){Bromley}, {Chen}, \&
  {Miller}}]{1997ApJ...475...57B}
{Bromley}, B.~C., {Chen}, K., \& {Miller}, W.~A. 1997, \apj, 475, 57

\bibitem[{{Bromley} {et~al.}(1998){Bromley}, {Miller}, \&
  {Pariev}}]{1998Natur.391...54B}
{Bromley}, B.~C., {Miller}, W.~A., \& {Pariev}, V.~I. 1998, \nat, 391, 54

\bibitem[{{Carter}(1968)}]{1968PhRv..174.1559C}
{Carter}, B. 1968, Physical Review, 174, 1559

\bibitem[{{Chandrasekhar}(1983)}]{1983mtbh.book.....C}
{Chandrasekhar}, S. 1983, {The mathematical theory of black holes}, ed.
  {Chandrasekhar, S.}

\bibitem[{{Demianski} \& {Ivanov}(1997)}]{1997A&A...324..829D}
{Demianski}, M., \& {Ivanov}, P.~B. 1997, \aap, 324, 829

\bibitem[{{Dexter} \& {Agol}(2009)}]{2009ApJ...696.1616D}
{Dexter}, J., \& {Agol}, E. 2009, \apj, 696, 1616

\bibitem[{{Doeleman} {et~al.}(2009){Doeleman}, {Agol}, {Backer}, {Baganoff},
  {Bower}, {Broderick}, {Fabian}, {Fish}, {Gammie}, {Ho}, {Honman},
  {Krichbaum}, {Loeb}, {Marrone}, {Reid}, {Rogers}, {Shapiro}, {Strittmatter},
  {Tilanus}, {Weintroub}, {Whitney}, {Wright}, \&
  {Ziurys}}]{2009astro2010S..68D}
{Doeleman}, S., {Agol}, E., {Backer}, D., {Baganoff}, F., {Bower}, G.~C.,
  {Broderick}, A., {Fabian}, A., {Fish}, V., {Gammie}, C., {Ho}, P., {Honman},
  M., {Krichbaum}, T., {Loeb}, A., {Marrone}, D., {Reid}, M., {Rogers}, A.,
  {Shapiro}, I., {Strittmatter}, P., {Tilanus}, R., {Weintroub}, J., {Whitney},
  A., {Wright}, M., \& {Ziurys}, L. 2009, in ArXiv Astrophysics e-prints, Vol.
  2010, astro2010: The Astronomy and Astrophysics Decadal Survey, 68--+

\bibitem[{{Fabian} {et~al.}(1989){Fabian}, {Rees}, {Stella}, \&
  {White}}]{1989MNRAS.238..729F}
{Fabian}, A.~C., {Rees}, M.~J., {Stella}, L., \& {White}, N.~E. 1989, \mnras,
  238, 729

\bibitem[{{Fanton} {et~al.}(1997){Fanton}, {Calvani}, {de Felice}, \&
  {Cadez}}]{1997PASJ...49..159F}
{Fanton}, C., {Calvani}, M., {de Felice}, F., \& {Cadez}, A. 1997, \pasj, 49,
  159

\bibitem[{{Frank} {et~al.}(2002){Frank}, {King}, \& {Raine}}]{Frankbook2002}
{Frank}, J., {King}, A., \& {Raine}, D.~J. 2002, {Accretion Power in
  Astrophysics: Third Edition}

\bibitem[{{Fuerst} \& {Wu}(2007)}]{2007A&A...474...55F}
{Fuerst}, S.~V., \& {Wu}, K. 2007, \aap, 474, 55

\bibitem[{{Hartnoll} \& {Blackman}(2000)}]{2000MNRAS.317..880H}
{Hartnoll}, S.~A., \& {Blackman}, E.~G. 2000, \mnras, 317, 880

\bibitem[{{Hartnoll} \& {Blackman}(2002)}]{2002MNRAS.332L...1H}
---. 2002, \mnras, 332, L1

\bibitem[{{Ivanov} \& {Illarionov}(1997)}]{1997MNRAS.285..394I}
{Ivanov}, P.~B., \& {Illarionov}, A.~F. 1997, \mnras, 285, 394

\bibitem[{{Karas} {et~al.}(2001){Karas}, {Martocchia}, \&
  {Subr}}]{2001PASJ...53..189K}
{Karas}, V., {Martocchia}, A., \& {Subr}, L. 2001, \pasj, 53, 189

\bibitem[{{Katz}(1973)}]{1973Natur.246...87K}
{Katz}, J.~I. 1973, \nat, 246, 87

\bibitem[{{Kinney} {et~al.}(2000){Kinney}, {Schmitt}, {Clarke}, {Pringle},
  {Ulvestad}, \& {Antonucci}}]{2000ApJ...537..152K}
{Kinney}, A.~L., {Schmitt}, H.~R., {Clarke}, C.~J., {Pringle}, J.~E.,
  {Ulvestad}, J.~S., \& {Antonucci}, R.~R.~J. 2000, \apj, 537, 152

\bibitem[{{Laor}(1991)}]{1991ApJ...376...90L}
{Laor}, A. 1991, \apj, 376, 90

\bibitem[{{Larwood} {et~al.}(1996){Larwood}, {Nelson}, {Papaloizou}, \&
  {Terquem}}]{1996MNRAS.282..597L}
{Larwood}, J.~D., {Nelson}, R.~P., {Papaloizou}, J.~C.~B., \& {Terquem}, C.
  1996, \mnras, 282, 597

\bibitem[{{Lubow} {et~al.}(2002){Lubow}, {Ogilvie}, \&
  {Pringle}}]{2002MNRAS.337..706L}
{Lubow}, S.~H., {Ogilvie}, G.~I., \& {Pringle}, J.~E. 2002, \mnras, 337, 706

\bibitem[{{Maloney} {et~al.}(1998){Maloney}, {Begelman}, \&
  {Nowak}}]{1998ApJ...504...77M}
{Maloney}, P.~R., {Begelman}, M.~C., \& {Nowak}, M.~A. 1998, \apj, 504, 77

\bibitem[{{Maloney} {et~al.}(1996){Maloney}, {Begelman}, \&
  {Pringle}}]{1996ApJ...472..582M}
{Maloney}, P.~R., {Begelman}, M.~C., \& {Pringle}, J.~E. 1996, \apj, 472, 582

\bibitem[{{Margon}(1984)}]{1984ARA&A..22..507M}
{Margon}, B. 1984, \araa, 22, 507

\bibitem[{{Martin}(2008)}]{2008MNRAS.387..830M}
{Martin}, R.~G. 2008, \mnras, 387, 830

\bibitem[{{Matt} {et~al.}(1991){Matt}, {Perola}, \&
  {Piro}}]{1991A&A...247...25M}
{Matt}, G., {Perola}, G.~C., \& {Piro}, L. 1991, \aap, 247, 25

\bibitem[{{McClintock} {et~al.}(2011){McClintock}, {Narayan}, {Davis}, {Gou},
  {Kulkarni}, {Orosz}, {Penna}, {Remillard}, \&
  {Steiner}}]{2011arXiv1101.0811M}
{McClintock}, J.~E., {Narayan}, R., {Davis}, S.~W., {Gou}, L., {Kulkarni}, A.,
  {Orosz}, J.~A., {Penna}, R.~F., {Remillard}, R.~A., \& {Steiner}, J.~F. 2011,
  ArXiv e-prints

\bibitem[{{Melia}(2007)}]{2007gsbh.book.....M}
{Melia}, F. 2007, {The Galactic Supermassive Black Hole} (Princeton University
  Press)

\bibitem[{{Miller}(2007)}]{2007ARA&A..45..441M}
{Miller}, J.~M. 2007, \araa, 45, 441

\bibitem[{{Miller} {et~al.}(2002{\natexlab{a}}){Miller}, {Fabian}, {Wijnands},
  {Remillard}, {Wojdowski}, {Schulz}, {Di Matteo}, {Marshall}, {Canizares},
  {Pooley}, \& {Lewin}}]{2002ApJ...578..348M}
{Miller}, J.~M., {Fabian}, A.~C., {Wijnands}, R., {Remillard}, R.~A.,
  {Wojdowski}, P., {Schulz}, N.~S., {Di Matteo}, T., {Marshall}, H.~L.,
  {Canizares}, C.~R., {Pooley}, D., \& {Lewin}, W.~H.~G. 2002{\natexlab{a}},
  \apj, 578, 348

\bibitem[{{Miller} {et~al.}(2002{\natexlab{b}}){Miller}, {Fabian}, {Wijnands},
  {Reynolds}, {Ehle}, {Freyberg}, {van der Klis}, {Lewin}, {Sanchez-Fernandez},
  \& {Castro-Tirado}}]{2002ApJ...570L..69M}
{Miller}, J.~M., {Fabian}, A.~C., {Wijnands}, R., {Reynolds}, C.~S., {Ehle},
  M., {Freyberg}, M.~J., {van der Klis}, M., {Lewin}, W.~H.~G.,
  {Sanchez-Fernandez}, C., \& {Castro-Tirado}, A.~J. 2002{\natexlab{b}}, \apjl,
  570, L69

\bibitem[{{Miniutti} {et~al.}(2007){Miniutti}, {Fabian}, {Anabuki}, {Crummy},
  {Fukazawa}, {Gallo}, {Haba}, {Hayashida}, {Holt}, {Kunieda}, {Larsson},
  {Markowitz}, {Matsumoto}, {Ohno}, {Reeves}, {Takahashi}, {Tanaka},
  {Terashima}, {Torii}, {Ueda}, {Ushio}, {Watanabe}, {Yamauchi}, \&
  {Yaqoob}}]{2007PASJ...59S.315M}
{Miniutti}, G., {Fabian}, A.~C., {Anabuki}, N., {Crummy}, J., {Fukazawa}, Y.,
  {Gallo}, L., {Haba}, Y., {Hayashida}, K., {Holt}, S., {Kunieda}, H.,
  {Larsson}, J., {Markowitz}, A., {Matsumoto}, C., {Ohno}, M., {Reeves}, J.~N.,
  {Takahashi}, T., {Tanaka}, Y., {Terashima}, Y., {Torii}, K., {Ueda}, Y.,
  {Ushio}, M., {Watanabe}, S., {Yamauchi}, M., \& {Yaqoob}, T. 2007, \pasj, 59,
  315

\bibitem[{{Miniutti} {et~al.}(2004){Miniutti}, {Fabian}, \&
  {Miller}}]{2004MNRAS.351..466M}
{Miniutti}, G., {Fabian}, A.~C., \& {Miller}, J.~M. 2004, \mnras, 351, 466

\bibitem[{{Misner} {et~al.}(1973){Misner}, {Thorne}, \& {Wheeler}}]{MTW}
{Misner}, C.~W., {Thorne}, K.~S., \& {Wheeler}, J.~A. 1973, {Gravitation} (San
  Francisco: Freeman)

\bibitem[{{Miyoshi} {et~al.}(1995){Miyoshi}, {Moran}, {Herrnstein},
  {Greenhill}, {Nakai}, {Diamond}, \& {Inoue}}]{1995Natur.373..127M}
{Miyoshi}, M., {Moran}, J., {Herrnstein}, J., {Greenhill}, L., {Nakai}, N.,
  {Diamond}, P., \& {Inoue}, M. 1995, \nat, 373, 127

\bibitem[{{Murphy} {et~al.}(2009){Murphy}, {Yaqoob}, {Karas}, \& {Dov{\v
  c}iak}}]{2009ApJ...701..635M}
{Murphy}, K.~D., {Yaqoob}, T., {Karas}, V., \& {Dov{\v c}iak}, M. 2009, \apj,
  701, 635

\bibitem[{{Nandra} {et~al.}(1997){Nandra}, {George}, {Mushotzky}, {Turner}, \&
  {Yaqoob}}]{1997ApJ...477..602N}
{Nandra}, K., {George}, I.~M., {Mushotzky}, R.~F., {Turner}, T.~J., \&
  {Yaqoob}, T. 1997, \apj, 477, 602

\bibitem[{{Novikov} \& {Thorne}(1973)}]{1973blho.conf..343N}
{Novikov}, I.~D., \& {Thorne}, K.~S. 1973, in Black Holes (Les Astres Occlus),
  ed. {C.~Dewitt \& B.~S.~Dewitt}, 343--450

\bibitem[{{Page} \& {Thorne}(1974)}]{1974ApJ...191..499P}
{Page}, D.~N., \& {Thorne}, K.~S. 1974, \apj, 191, 499

\bibitem[{{Press} {et~al.}(1992){Press}, {Teukolsky}, {Vetterling}, \&
  {Flannery}}]{1992nrfa.book.....P}
{Press}, W.~H., {Teukolsky}, S.~A., {Vetterling}, W.~T., \& {Flannery}, B.~P.
  1992, {Numerical recipes in FORTRAN. The art of scientific computing}, ed.
  {Press, W.~H., Teukolsky, S.~A., Vetterling, W.~T., \& Flannery, B.~P. }

\bibitem[{{Pringle}(1996)}]{1996MNRAS.281..357P}
{Pringle}, J.~E. 1996, \mnras, 281, 357

\bibitem[{{Rauch} \& {Blandford}(1994)}]{1994ApJ...421...46R}
{Rauch}, K.~P., \& {Blandford}, R.~D. 1994, \apj, 421, 46

\bibitem[{{Reynolds} \& {Nowak}(2003)}]{2003PhR...377..389R}
{Reynolds}, C.~S., \& {Nowak}, M.~A. 2003, \physrep, 377, 389

\bibitem[{{Rossi} {et~al.}(2005){Rossi}, {Homan}, {Miller}, \&
  {Belloni}}]{2005MNRAS.360..763R}
{Rossi}, S., {Homan}, J., {Miller}, J.~M., \& {Belloni}, T. 2005, \mnras, 360,
  763

\bibitem[{{Schmitt} {et~al.}(2002){Schmitt}, {Pringle}, {Clarke}, \&
  {Kinney}}]{2002ApJ...575..150S}
{Schmitt}, H.~R., {Pringle}, J.~E., {Clarke}, C.~J., \& {Kinney}, A.~L. 2002,
  \apj, 575, 150

\bibitem[{{Shakura} \& {Sunyaev}(1973)}]{1973A&A....24..337S}
{Shakura}, N.~I., \& {Sunyaev}, R.~A. 1973, \aap, 24, 337

\bibitem[{{Sochora} {et~al.}(2011){Sochora}, {Karas}, {Svoboda}, \& {Dov{\v
  c}iak}}]{2011MNRAS.tmp.1522S}
{Sochora}, V., {Karas}, V., {Svoboda}, J., \& {Dov{\v c}iak}, M. 2011, \mnras,
  1522

\bibitem[{{Tanaka} {et~al.}(1995){Tanaka}, {Nandra}, {Fabian}, {Inoue},
  {Otani}, {Dotani}, {Hayashida}, {Iwasawa}, {Kii}, {Kunieda}, {Makino}, \&
  {Matsuoka}}]{1995Natur.375..659T}
{Tanaka}, Y., {Nandra}, K., {Fabian}, A.~C., {Inoue}, H., {Otani}, C.,
  {Dotani}, T., {Hayashida}, K., {Iwasawa}, K., {Kii}, T., {Kunieda}, H.,
  {Makino}, F., \& {Matsuoka}, M. 1995, \nat, 375, 659

\bibitem[{{Tananbaum} {et~al.}(1972){Tananbaum}, {Gursky}, {Kellogg},
  {Levinson}, {Schreier}, \& {Giacconi}}]{1972ApJ...174L.143T}
{Tananbaum}, H., {Gursky}, H., {Kellogg}, E.~M., {Levinson}, R., {Schreier},
  E., \& {Giacconi}, R. 1972, \apjl, 174, L143+

\bibitem[{{Terquem} \& {Bertout}(1993)}]{1993A&A...274..291T}
{Terquem}, C., \& {Bertout}, C. 1993, \aap, 274, 291

\bibitem[{{Thorne}(1974)}]{1974ApJ...191..507T}
{Thorne}, K.~S. 1974, \apj, 191, 507

\bibitem[{{Turner} {et~al.}(1997){Turner}, {George}, {Nandra}, \&
  {Mushotzky}}]{1997ApJ...488..164T}
{Turner}, T.~J., {George}, I.~M., {Nandra}, K., \& {Mushotzky}, R.~F. 1997,
  \apj, 488, 164

\bibitem[{{Wijers} \& {Pringle}(1999)}]{1999MNRAS.308..207W}
{Wijers}, R.~A.~M.~J., \& {Pringle}, J.~E. 1999, \mnras, 308, 207

\bibitem[{{Wilms} {et~al.}(2001){Wilms}, {Reynolds}, {Begelman}, {Reeves},
  {Molendi}, {Staubert}, \& {Kendziorra}}]{2001MNRAS.328L..27W}
{Wilms}, J., {Reynolds}, C.~S., {Begelman}, M.~C., {Reeves}, J., {Molendi}, S.,
  {Staubert}, R., \& {Kendziorra}, E. 2001, \mnras, 328, L27

\bibitem[{{Wu} {et~al.}(2008){Wu}, {Wang}, \& {Dong}}]{2008MNRAS.389..213W}
{Wu}, S., {Wang}, T., \& {Dong}, X. 2008, \mnras, 389, 213

\bibitem[{{Wu} \& {Wang}(2007)}]{2007MNRAS.378..841W}
{Wu}, S.-M., \& {Wang}, T.-G. 2007, \mnras, 378, 841

\bibitem[{{Young} {et~al.}(1998){Young}, {Ross}, \&
  {Fabian}}]{1998MNRAS.300L..11Y}
{Young}, A.~J., {Ross}, R.~R., \& {Fabian}, A.~C. 1998, \mnras, 300, L11

\end{thebibliography}

\end{document}